\documentclass[10pt,a4paper]{article}

\usepackage{jheppub}

\long\def\symbolfootnote[#1]#2{\begingroup%
\def\thefootnote{\fnsymbol{footnote}}\footnote[#1]{#2}\endgroup}

\def\hermesauthor[#1]#2{{#2}$^{\, #1}$}
\def\hermesinstitute[#1]#2{$^{#1\,}$ {#2}\\}

\renewcommand{\thefootnote}{\alph{footnote}}
\def\nowat[#1]#2{\(^,\)\footnote[#1]{#2}}

\def\amp2{{\cal T}}

\newcommand{\etal}{\emph{et al.}}

\preprint{ \today, version 4.2 (after JHEP referee)}

\title{Beam-helicity asymmetry arising from deeply virtual Compton scattering measured with kinematically complete event reconstruction}

\collaboration{The {\sc Hermes} Collaboration}
\affiliation{DESY -- HERMES, Notkestra\ss e 85, D-22607 Hamburg}
\emailAdd{management@hermes.desy.de}

\abstract{
The beam-helicity asymmetry in exclusive electroproduction of real photons by the longitudinally polarized {\sc Hera} positron beam scattering off an unpolarized hydrogen target is measured at {\sc Hermes}. The asymmetry arises from deeply virtual Compton scattering and its interference with the Bethe--Heitler process. Azimuthal amplitudes of the beam-helicity asymmetry are extracted from a data sample consisting of $ep\rightarrow ep\gamma$ events with detection of all particles in the final state including the recoiling proton. 
 The installation of a recoil detector, while reducing the acceptance of the experiment, allows the elimination of background from $ep\rightarrow eN\pi\gamma$ events, which was estimated to contribute an average of about 12\%  to the signal in previous {\sc Hermes} publications. 
The removal of this background from the present data sample is shown to increase the magnitude of the leading asymmetry amplitude by $0.054\pm0.016$ to $-0.328\pm0.027\;\mathrm{(stat.)}\pm0.045\;\mathrm{(syst.)}$.}

\keywords{lepton-nucleon scattering}

\notoc

\begin{document}

\maketitle\flushbottom\newpage

\pagestyle{myplain}
\pagenumbering{roman}

\section*{The {\sc Hermes} Collaboration}

{%
\begin{flushleft} 
\bf
\hermesauthor[13,16]{A.~Airapetian},
\hermesauthor[27]{N.~Akopov},
\hermesauthor[6]{Z.~Akopov},
\hermesauthor[7]{E.C.~Aschenauer}\nowat[1]{Now at: Brookhaven National Laboratory, Upton, New York 11772-5000, USA},
\hermesauthor[26]{W.~Augustyniak},
\hermesauthor[27]{R.~Avakian},
\hermesauthor[27]{A.~Avetissian},
\hermesauthor[6]{E.~Avetisyan},
\hermesauthor[19]{S.~Belostotski},
\hermesauthor[18,25]{H.P.~Blok},
\hermesauthor[6]{A.~Borissov},
\hermesauthor[14]{J.~Bowles},
\hermesauthor[13]{I.~Brodski},
\hermesauthor[20]{V.~Bryzgalov},
\hermesauthor[14]{J.~Burns},
\hermesauthor[10]{M.~Capiluppi},
\hermesauthor[11]{G.P.~Capitani},
\hermesauthor[22]{E.~Cisbani},
\hermesauthor[10]{G.~Ciullo},
\hermesauthor[10]{M.~Contalbrigo},
\hermesauthor[10]{P.F.~Dalpiaz},
\hermesauthor[6]{W.~Deconinck},
\hermesauthor[2]{R.~De~Leo},
\hermesauthor[12,6,23]{L.~De~Nardo},
\hermesauthor[11]{E.~De~Sanctis},
\hermesauthor[15,9]{M.~Diefenthaler},
\hermesauthor[11]{P.~Di~Nezza},
\hermesauthor[13]{M.~D\"uren},
\hermesauthor[13]{M.~Ehrenfried},
\hermesauthor[27]{G.~Elbakian},
\hermesauthor[5]{F.~Ellinghaus},
\hermesauthor[13]{E.~Etzelm\"uller},
\hermesauthor[7]{R.~Fabbri},
\hermesauthor[11]{A.~Fantoni},
\hermesauthor[23]{L.~Felawka},
\hermesauthor[22]{S.~Frullani},
\hermesauthor[20]{G.~Gapienko},
\hermesauthor[20]{V.~Gapienko},
\hermesauthor[22]{F.~Garibaldi},
\hermesauthor[6,19,23]{G.~Gavrilov},
\hermesauthor[27]{V.~Gharibyan},
\hermesauthor[15,10]{F.~Giordano},
\hermesauthor[16]{S.~Gliske},
\hermesauthor[7]{M.~Golembiovskaya},
\hermesauthor[7]{I.M.~Gregor},
\hermesauthor[7]{H.~Guler},
\hermesauthor[6]{M.~Hartig},
\hermesauthor[11]{D.~Hasch},
\hermesauthor[7]{A.~Hillenbrand},
\hermesauthor[14]{M.~Hoek},
\hermesauthor[6]{Y.~Holler},
\hermesauthor[7]{I.~Hristova},
\hermesauthor[20]{A.~Ivanilov},
\hermesauthor[1]{H.E.~Jackson},
\hermesauthor[12]{H.S.~Jo},
\hermesauthor[15]{S.~Joosten},
\hermesauthor[14]{R.~Kaiser}\nowat[2]{Present address: International Atomic Energy Agency, A-1400 Vienna, Austria},
\hermesauthor[27]{G.~Karyan},
\hermesauthor[14,13]{T.~Keri},
\hermesauthor[5]{E.~Kinney},
\hermesauthor[19]{A.~Kisselev},
\hermesauthor[20]{V.~Korotkov},
\hermesauthor[17]{V.~Kozlov},
\hermesauthor[9]{B.~Krauss},
\hermesauthor[9,19]{P.~Kravchenko},
\hermesauthor[8]{V.G.~Krivokhijine},
\hermesauthor[2]{L.~Lagamba},
\hermesauthor[18]{L.~Lapik\'as},
\hermesauthor[14]{I.~Lehmann},
\hermesauthor[10]{P.~Lenisa},
\hermesauthor[12]{A.~L\'opez~Ruiz},
\hermesauthor[16]{W.~Lorenzon},
\hermesauthor[13]{S.~Lu},
\hermesauthor[7]{X.~Lu},
\hermesauthor[3]{B.-Q.~Ma},
\hermesauthor[14]{D.~Mahon},
\hermesauthor[15]{N.C.R.~Makins},
\hermesauthor[19]{S.I.~Manaenkov},
\hermesauthor[22]{L.~Manfr\'e},
\hermesauthor[3]{Y.~Mao},
\hermesauthor[26]{B.~Marianski},
\hermesauthor[6,5]{A.~Martinez de la Ossa},
\hermesauthor[27]{H.~Marukyan},
\hermesauthor[23]{C.A.~Miller},
\hermesauthor[24]{Y.~Miyachi}\nowat[3]{Now at: Department of Physics, Yamagata University, Yamagata 990-8560, Japan},
\hermesauthor[27]{A.~Movsisyan},
\hermesauthor[14]{M.~Murray},
\hermesauthor[6,9]{A.~Mussgiller},
\hermesauthor[2]{E.~Nappi},
\hermesauthor[19]{Y.~Naryshkin},
\hermesauthor[9]{A.~Nass},
\hermesauthor[7]{M.~Negodaev},
\hermesauthor[7]{W.-D.~Nowak},
\hermesauthor[14]{A.~Osborne},
\hermesauthor[10]{L.L.~Pappalardo},
\hermesauthor[13]{R.~Perez-Benito},
\hermesauthor[27]{A.~Petrosyan},
\hermesauthor[1]{P.E.~Reimer},
\hermesauthor[11]{A.R.~Reolon},
\hermesauthor[7]{C.~Riedl},
\hermesauthor[9]{K.~Rith},
\hermesauthor[14]{G.~Rosner},
\hermesauthor[6]{A.~Rostomyan},
\hermesauthor[13]{L.~Rubacek},
\hermesauthor[1,15]{J.~Rubin},
\hermesauthor[12]{D.~Ryckbosch},
\hermesauthor[21]{A.~Sch\"afer},
\hermesauthor[4,12]{G.~Schnell},
\hermesauthor[6]{K.P.~Sch\"uler},
\hermesauthor[14]{B.~Seitz},
\hermesauthor[14]{C.~Shearer},
\hermesauthor[24]{T.-A.~Shibata},
\hermesauthor[8]{V.~Shutov},
\hermesauthor[10]{M.~Stancari},
\hermesauthor[10]{M.~Statera},
\hermesauthor[18]{J.J.M.~Steijger},
\hermesauthor[7]{J.~Stewart},
\hermesauthor[27]{S.~Taroian},
\hermesauthor[17]{A.~Terkulov},
\hermesauthor[15]{R.~Truty},
\hermesauthor[26]{A.~Trzcinski},
\hermesauthor[12]{M.~Tytgat},
\hermesauthor[12]{Y.~Van~Haarlem},
\hermesauthor[4,12]{C.~Van~Hulse}, 
\hermesauthor[19]{D.~Veretennikov},
\hermesauthor[19]{V.~Vikhrov},
\hermesauthor[2]{I.~Vilardi},
\hermesauthor[3]{S.~Wang},
\hermesauthor[7,9]{S.~Yaschenko},
\hermesauthor[6]{Z.~Ye},
\hermesauthor[23]{S.~Yen},
\hermesauthor[6,13]{V.~Zagrebelnyy},
\hermesauthor[9]{D.~Zeiler},
\hermesauthor[6]{B.~Zihlmann},
\hermesauthor[26]{P.~Zupranski}
\end{flushleft} 
}
\bigskip
{\it
\begin{flushleft} 
\hermesinstitute[1]{Physics Division, Argonne National Laboratory, Argonne, Illinois 60439-4843, USA}
\hermesinstitute[2]{Istituto Nazionale di Fisica Nucleare, Sezione di Bari, 70124 Bari, Italy}
\hermesinstitute[3]{School of Physics, Peking University, Beijing 100871, China}
\hermesinstitute[4]{Department of Theoretical Physics, University of the Basque Country UPV/EHU, 48080 Bilbao, Spain and IKERBASQUE, Basque Foundation for Science, 48011 Bilbao, Spain}
\hermesinstitute[5]{Nuclear Physics Laboratory, University of Colorado, Boulder, Colorado 80309-0390, USA}
\hermesinstitute[6]{DESY, 22603 Hamburg, Germany}
\hermesinstitute[7]{DESY, 15738 Zeuthen, Germany}
\hermesinstitute[8]{Joint Institute for Nuclear Research, 141980 Dubna, Russia}
\hermesinstitute[9]{Physikalisches Institut, Universit\"at Erlangen-N\"urnberg, 91058 Erlangen, Germany}
\hermesinstitute[10]{Istituto Nazionale di Fisica Nucleare, Sezione di Ferrara and Dipartimento di Fisica, Universit\`a di Ferrara, 44100 Ferrara, Italy}
\hermesinstitute[11]{Istituto Nazionale di Fisica Nucleare, Laboratori Nazionali di Frascati, 00044 Frascati, Italy}
\hermesinstitute[12]{Department of Physics and Astronomy, Ghent University, 9000 Gent, Belgium}
\hermesinstitute[13]{Physikalisches Institut, Universit\"at Gie{\ss}en, 35392 Gie{\ss}en, Germany}
\hermesinstitute[14]{SUPA, School of Physics and Astronomy, University of Glasgow, Glasgow G12 8QQ, United Kingdom}
\hermesinstitute[15]{Department of Physics, University of Illinois, Urbana, Illinois 61801-3080, USA}
\hermesinstitute[16]{Randall Laboratory of Physics, University of Michigan, Ann Arbor, Michigan 48109-1040, USA }
\hermesinstitute[17]{Lebedev Physical Institute, 117924 Moscow, Russia}
\hermesinstitute[18]{National Institute for Subatomic Physics (Nikhef), 1009 DB Amsterdam, The Netherlands}
\hermesinstitute[19]{B.P. Konstantinov Petersburg Nuclear Physics Institute, Gatchina, 188300 Leningrad Region, Russia}
\hermesinstitute[20]{Institute for High Energy Physics, Protvino, 142281 Moscow Region, Russia}
\hermesinstitute[21]{Institut f\"ur Theoretische Physik, Universit\"at Regensburg, 93040 Regensburg, Germany}
\hermesinstitute[22]{Istituto Nazionale di Fisica Nucleare, Sezione di Roma, Gruppo Collegato Sanit\`a and Istituto Superiore di Sanit\`a, 00161 Roma, Italy}
\hermesinstitute[23]{TRIUMF, Vancouver, British Columbia V6T 2A3, Canada}
\hermesinstitute[24]{Department of Physics, Tokyo Institute of Technology, Tokyo 152, Japan}
\hermesinstitute[25]{Department of Physics and Astronomy, VU University, 1081 HV Amsterdam, The Netherlands}
\hermesinstitute[26]{National Centre for Nuclear Research, 00-689 Warsaw, Poland}
\hermesinstitute[27]{Yerevan Physics Institute, 375036 Yerevan, Armenia}
\end{flushleft} 
}

\clearpage
\pagenumbering{arabic}

\noindent\rule\textwidth{.1pt}
\tableofcontents
\afterTocSpace
\noindent\rule\textwidth{.1pt}

\pagestyle{myplain}

\section{Introduction}
\label{sec:Introduction}

Generalized Parton Distributions (GPDs) \cite{Mueller, Radyushkin, Ji}  describe the soft, non-perturbative part of hard exclusive reactions, e.g., hard exclusive leptoproduction of a real photon or a meson leaving the nucleon intact (possibly modulo isospin rotation). 
These distributions have quickly increased in importance in QCD spin physics since it was shown that they can provide access to the total angular momentum carried by quarks in the nucleon (and also by gluons, in principle) \cite{Ji_2}. The resulting intense theoretical activity is exemplified by the demonstration that GPDs can be considered as form factors, dissected in longitudinal nucleon momentum, describing transverse density distributions of quarks (and gluons) \cite{Burkardt} (``nucleon tomography'').
Hard exclusive processes provide experimental access to GPDs.
One of the experimental challenges in these measurements is the selection of truly exclusive final states, discriminating against the excitation of baryonic resonances. One solution to this problem is the detection of all particles in the final state with adequate kinematic resolution. 

Generalized parton distributions depend on four kinematic variables: $t$, $x$, $\xi$, and $Q^2$. 
The Mandelstam variable $t=(p-p^{\prime})^2$ is the square of the difference between the initial ($p$) and final ($p^{\prime}$) four-momenta of the target proton. The variable $x$ is the average of the initial and final fractions of the (large) target longitudinal momentum that is carried by the struck parton, and the variable $\xi$, known as the skewness, is half of the difference between these fractions. 
The evolution of GPDs with the photon virtuality $Q^2\equiv -q^2$ is analogous to that of parton distribution functions, with $q=k-k^{\prime}$ being the difference between the  four-momenta of the incident and the scattered leptons. 
Currently, there exist no hard exclusive measurements that provide access to $x$.
Because of the lack of consensus about how to define $\xi$ in terms of experimental observables, the results are typically reported by {\sc Hermes} as projections in $x_{\textrm{B}}\equiv Q^2/(2pq)$, to which $\xi$ can be related through $\xi\simeq x_{\textrm{B}}/(2-x_{\textrm{B}})$ in the generalized Bjorken limit of large $Q^2$ and fixed $x_{\textrm{B}}$ and $t$.

Several GPDs describe various possible helicity transitions of the struck quark and/or the nucleon as a whole. At leading twist (i.e., twist-2) and for a spin-1/2 target such as the proton, four chiral-even GPDs ($H$, $\widetilde{H}$, $E$, $\widetilde{E}$) are required to describe processes that conserve the helicity of the struck quark.
Deeply Virtual Compton Scattering (DVCS), i.e., the hard exclusive leptoproduction of a real photon, has the most reliable interpretation in terms of GPDs among all presently practical hard exclusive probes.
The measurement of the DVCS process on unpolarized protons is most sensitive to GPD $H$, which describes the transition that conserves the helicities of both the struck quark and the nucleon.
Such measurements were performed by {\sc Hermes} \cite{Air01, Air06, PublicationDraft69, PublicationDraft90}, H1 \cite{h101, h105, h107, h109}, and {\sc Zeus} \cite{zeu03, zeu08} at {\sc Hera}, by the Hall A Collaboration \cite{Cam06} and by {\sc Clas} \cite{Ste01, Gir08, Gav09} at Jefferson Lab.

This paper reports a kinematically complete measurement of DVCS for a polarized lepton beam on an unpolarized hydrogen target with detection of all particles in the final state, including the recoil proton. It is the first measurement of this kind reported by {\sc Hermes}. The results of this measurement are compared to measurements without detection of the recoil proton, while accounting for the difference in acceptance in the two techniques.
Measurements without recoil-proton detection have also been reported in previous publications.

\section{Constraining GPDs through DVCS}
\label{sec:dvcs}

\begin{figure}[t]
\centerline{
\epsfig{file=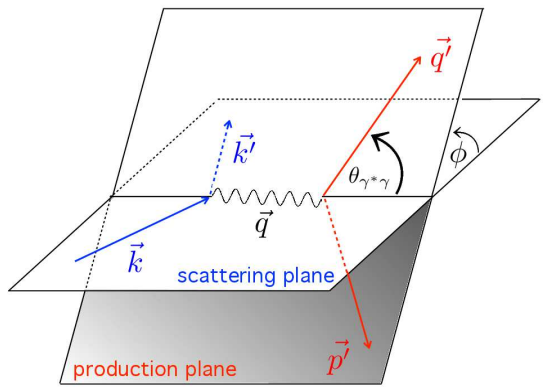, width=0.5\textwidth}
}
\caption{
\label{fig:phiangle} 
Momenta and azimuthal angle for exclusive electroproduction of real photons in the target rest frame. The quantity $\phi$ denotes the angle between the lepton scattering plane containing the three-momenta $\vec{k}$ and $\vec{k^\prime}$ of the incoming and outgoing lepton, respectively, and the plane correspondingly defined by $\vec{q}=\vec{k}-\vec{k^\prime}$ and the three-momentum $\vec{q^\prime}$ of the real photon.  Also indicated are the polar angle $\theta_{\gamma^*\gamma}$ between the three-momenta of the real ($\vec{q^\prime}$) and virtual ($\vec{q}$) photons, and the three-momentum of the recoil proton ($\vec{p^\prime}$).
}
\end{figure}

The four-fold differential cross section for exclusive single-photon production, $ep\rightarrow ep\gamma$, on an unpolarized proton target is given by \cite{DVCS2}

\begin{equation}
\frac{\textrm{d}^4\sigma}{ \textrm{d}Q^2\,\textrm{d}x_{\textrm{B}}\,\textrm{d} t\,\textrm{d}\phi}
=
\frac
{x_{\textrm{B}}e^6}
{32(2\pi)^4Q^4\sqrt{1+\epsilon^2}}
\lvert\mathcal{T}_{ep\rightarrow ep\gamma}\rvert^2,
\label{eq:xsec}
\end{equation}

\noindent where $\mathcal{T}_{ep\rightarrow ep\gamma}$ is the scattering amplitude for this process.
Here, $e$ is the elementary charge and $\epsilon=2x_{\textrm{B}}\frac{M_p}{Q}$ with $M_p$ the proton mass. The angle $\phi$ denotes the azimuthal orientation of the photon production plane with respect to the lepton scattering plane, as indicated in figure \ref{fig:phiangle}. This definition follows the Trento conventions \cite{Tre04}.

\begin{figure}[t]
\centerline{
\epsfig{file=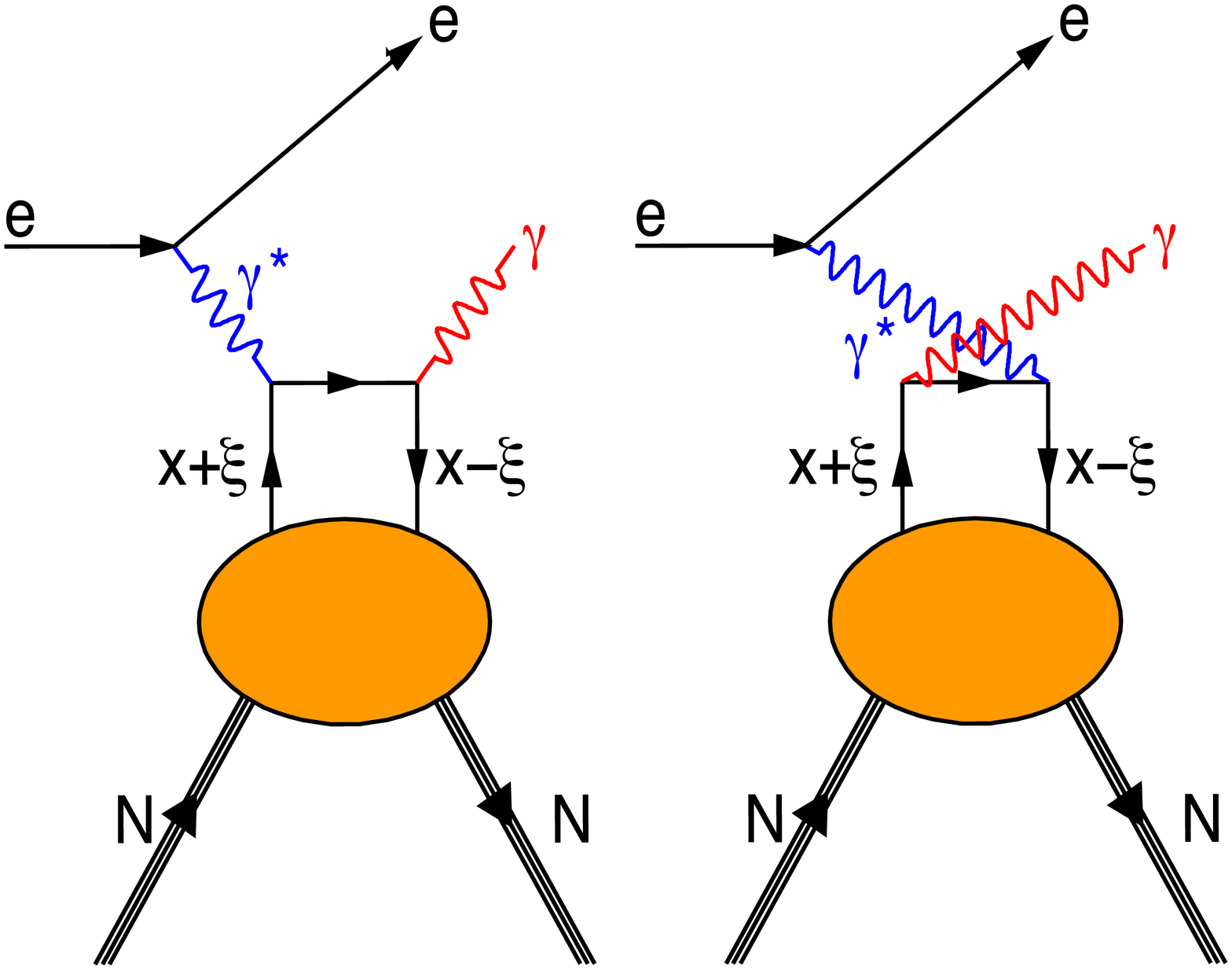, width=0.4\textwidth, angle=90}
\epsfig{file=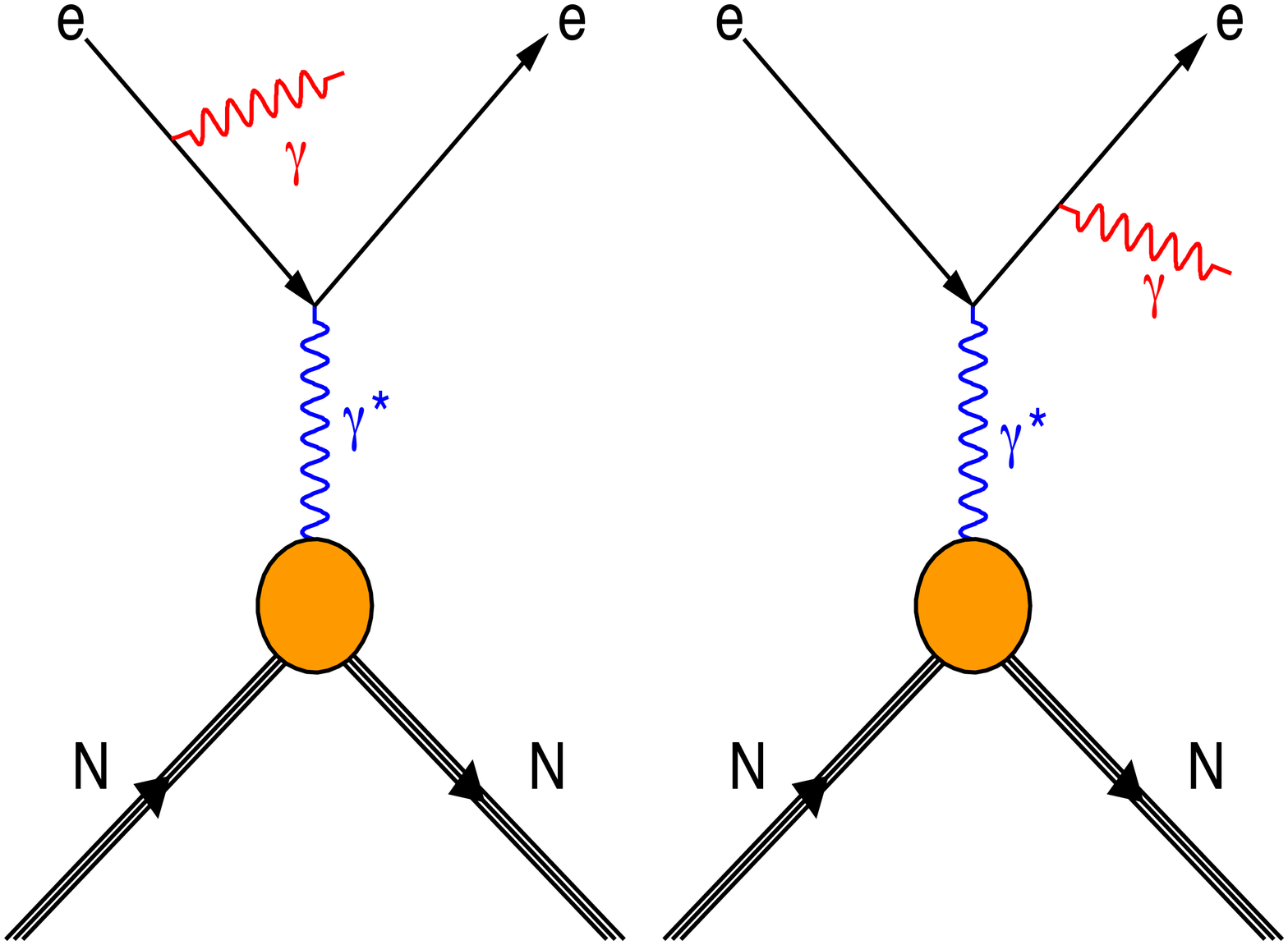, width=0.4\textwidth, angle=90}
}
\centerline{(a) DVCS \hspace{0.35\textwidth} (b) BH}
\caption{
\label{fig:dvcsbh} 
Leading-order diagrams for the channel $ep\rightarrow ep\gamma$ for (a) deeply virtual Compton scattering (DVCS) and (b) Bethe--Heitler (BH) processes. 
}
\end{figure}

In DVCS, a real photon is radiated from the struck parton, see figure \ref{fig:dvcsbh}(a). There is another process that contributes to the channel $ep\rightarrow ep\gamma$: the Bethe--Heitler (BH) process, where a bremsstrahlung photon is radiated from the incident or scattered lepton, see figure \ref{fig:dvcsbh}(b). The DVCS and BH processes have the same initial and final states, and therefore their scattering amplitudes interfere:
                                                                
\begin{equation}
\lvert\mathcal{T}_{ep\rightarrow ep\gamma}\rvert^2 = |\mathcal{T}_{\textrm{BH}}|^2 +|\mathcal{T}_{\textrm{DVCS}}|^2 + \mathcal{I},
\label{eq:scam}
\end{equation}

\noindent with the term $\mathcal{I}$ representing the interference between the scattering amplitudes of the BH process, $\mathcal{T}_{\mathrm{BH}}$, and the DVCS process, $\mathcal{T}_{\mathrm{DVCS}}$: 

\begin{equation}
\mathcal{I}=\mathcal{T}_{\textrm{BH}}\mathcal{T}_{\textrm{DVCS}}^{\ast}+\mathcal{T}_{\textrm{DVCS}}\mathcal{T}_{\textrm{BH}}^{\ast}.
\label{eq:intterm}
\end{equation}

\noindent 
The BH scattering amplitude is calculable in QED using the form factors of the proton measured in elastic scattering. 

The contributions to the cross section expressed through eqs.~(\ref{eq:xsec}) and (\ref{eq:scam}) can each be expanded in a harmonic series with respect to the angle $\phi$ \cite{DVCS2}:

\begin{eqnarray}
 |\mathcal{T}_{\textrm{BH}}|^2 &=&\frac{K_{\mathrm{BH}}}{\mathcal{P}_1(\phi)\mathcal{P}_2(\phi)}\left(c_0^{\mathrm{BH}}+\sum_{n=1}^2c_n^{\mathrm{BH}}\cos(n\phi)\right), \label{eq:taubh}\\
|\mathcal{T}_{\textrm{DVCS}}|^2&=&K_{\mathrm{DVCS}}\left(c_0^{\mathrm{DVCS}}+\sum_{n=1}^2c_n^{\mathrm{DVCS}}\cos(n\phi)+\lambda s_1^{\mathrm{DVCS}}\sin\phi\right), \label{eq:taudvcs}\\
 \mathcal{I}&=&\frac{-e_\ell K_{\mathcal{I}}}{\mathcal{P}_1(\phi)\mathcal{P}_2(\phi)}\left(c_0^{\mathcal{I}}+\sum_{n=1}^3c_n^{\mathcal{I}}\cos(n\phi)+\lambda\sum_{n=1}^2 s_n^{\mathcal{I}}\sin(n\phi)\right). \label{eq:tauint}
\end{eqnarray}

\noindent Here, $\mathcal{P}_1(\phi)$ and $\mathcal{P}_2(\phi)$ are the lepton propagators for the BH process, and $\lambda=\pm 1$ and $e_\ell=\pm 1$ are respectively the helicity and unit charge of the beam lepton. The quantities $K_{\mathrm{BH}}=1/(x_{\mathrm{B}}^2t(1+\epsilon^2))^2$, $K_{\mathrm{DVCS}}=1/Q^2$, and $K_{\mathcal{I}}=1/(y t x_{\mathrm{B}})$ are kinematic factors independent of $\phi$, with $y=(pq)/(pk)$. 
All beam-helicity dependent terms enter with sinusoidal harmonics due to parity conservation.

At the kinematic conditions of {\sc Hermes}, the square of the DVCS scattering amplitude yields only a small contribution to the $ep\rightarrow ep\gamma$ cross section, while the square of the BH scattering amplitude is much larger. 
Therefore, the contribution of the DVCS scattering amplitude to the cross section enters mainly through its interference with the BH scattering amplitude, thereby giving rise to cross-section asymmetries with respect to beam charge and beam helicity (and/or target polarization).
For a longitudinally polarized beam and an unpolarized target, the cross section of eq.~(\ref{eq:xsec}) can be expressed as:

\begin{equation}
\left.\frac{\textrm{d}^4\sigma}{ \textrm{d}Q^2\,\textrm{d}x_{\textrm{B}}\,\textrm{d} t\,\textrm{d}\phi}\right|_{\lambda,e_\ell}\equiv\sigma_{\mathrm{LU}}(\phi, e_\ell, \lambda)\equiv\sigma_{\mathrm{UU}}(\phi, e_\ell)\left[1+\lambda\mathcal{A}_{\mathrm{LU}}(\phi, e_\ell)\right],
\label{eq:sigma_lu}
\end{equation}

\noindent where $\sigma_{\mathrm{LU}}$ ($\sigma_{\mathrm{UU}}$) denotes the differential cross section for longitudinally polarized (unpolarized) beam and unpolarized target. On the right-hand side, the other kinematic dependences are omitted for brevity.
For an unpolarized beam, the cross section reads:

\begin{eqnarray}
\sigma_{\mathrm{UU}}(\phi, e_\ell) &=&
\frac
{x_{\textrm{B}}e_\ell^6}
{32(2\pi)^4Q^4\sqrt{1+\epsilon^2}}
\left[\frac{K_{\mathrm{BH}}\left(c_0^{\mathrm{BH}}+\sum_{n=1}^2c_n^{\mathrm{BH}}\cos(n\phi)\right)}{\mathcal{P}_1(\phi)\mathcal{P}_2(\phi)}\right.\nonumber\\
&&\left.
+K_{\mathrm{DVCS}}\left(c_0^{\mathrm{DVCS}}+\sum_{n=1}^2c_n^{\mathrm{DVCS}}\cos(n\phi)\right) -
\frac{e_\ell K_{\mathcal{I}}\left(c_0^{\mathcal{I}}+\sum_{n=1}^3c_n^{\mathcal{I}}\cos(n\phi)\right)}{\mathcal{P}_1(\phi)\mathcal{P}_2(\phi)}
\right]\!\!.
\label{eq:sigma_uu}
\end{eqnarray}

\noindent In eq.~(\ref{eq:sigma_lu}), $\mathcal{A}_{\mathrm{LU}}(\phi, e_\ell)$ denotes the single-charge beam-helicity asymmetry:

\begin{eqnarray}
\mathcal{A}_{\mathrm{LU}}(\phi, e_\ell)
&=&
\frac{\sigma_{\mathrm{LU}}(\phi, e_\ell,\lambda=+1)-\sigma_{\mathrm{LU}}(\phi, e_\ell,\lambda=-1)}{\sigma_{\mathrm{LU}}(\phi,e_\ell,\lambda=+1)+\sigma_{\mathrm{LU}}(\phi,e_\ell,\lambda=-1)}
\label{eq:alu_sbc1}\\
&=&
\frac{1}{\sigma_{\mathrm{UU}}(\phi, e_\ell)}\left[K_{\mathrm{DVCS}}\;s_1^{\mathrm{DVCS}}\sin\phi-e_\ell\frac{K_{\mathrm{I}}\sum_{n=1}^2 s_n^{\mathcal{I}}\sin(n\phi)}{\mathcal{P}_1(\phi)\mathcal{P}_2(\phi)}\right].
\label{eq:alu_sbc2}
\end{eqnarray}

\noindent It can be seen that $\mathcal{A}_{\mathrm{LU}}(\phi, e_\ell)$ is a mixture of contributions from beam-charge dependent interference and beam-charge independent squared DVCS terms in both its numerator and denominator, the latter being identified with $\sigma_{\mathrm{UU}}$. For data collected with only one beam charge, the two terms in eq.~(\ref{eq:alu_sbc2}) containing $s_1^{\mathrm{DVCS}}$ and $s_1^{\mathcal{I}}$ cannot be disentangled. However, at {\sc Hermes} kinematic conditions, the asymmetry is expected to be dominated by the term containing $s_1^{\mathcal{I}}$ \cite{DVCS2}.
Measurements that disentangle the contributions to the cross section from the interference term and the squared DVCS term of eq.~(\ref{eq:scam}) confirm this expectation by finding that the asymmetry related to the latter (and hence $s_1^{\mathrm{DVCS}}$) is negligible compared to that arising from the former \cite{PublicationDraft69}. 

The leading-twist Fourier coefficient $s_1^{\mathcal{I}}$ is related to GPDs via a complex function $\mathcal{C}^{\mathcal{I}}_{\mathrm{unp}}$:

\begin{equation}
s_1^{\mathcal{I}}\approx 8\frac{\sqrt{-t}}{Q}y(2-y)\;\Im\mathrm{m}(\mathcal{C}^{\mathcal{I}}_{\mathrm{unp}}).
\label{eq:s1ciunp}
\end{equation}

\noindent (The coefficient $s_2^{\mathcal{I}}$ enters at higher twist.)
The function $\mathcal{C}^{\mathcal{I}}_{\mathrm{unp}}$ is a linear combination of the Compton Form Factors (CFFs) $\mathcal{H}$, $\widetilde{\mathcal{H}}$, and $\mathcal{E}$ \cite{DVCS2}, which are flavor sums of convolutions of the corresponding GPDs
$H$, $\widetilde{H}$, and $E$ with hard scattering coefficient functions. This linear combination reads:

\begin{equation}
\mathcal{C}^{\mathcal{I}}_{\mathrm{unp}}=F_1\mathcal{H}+\frac{x_\mathrm{B}}{2-x_{\mathrm{B}}}(F_1+F_2)\widetilde{\mathcal{H}}-\frac{t}{4M_p^2}F_2\mathcal{E},
\label{eq:ciunp}
\end{equation}

\noindent where $F_1$ and $F_2$ are respectively the Dirac and Pauli form factors of the nucleon. At {\sc Hermes} kinematic conditions, where both $x_{\mathrm{B}}$ and ${-t}/{M_p^2}$ are of order 0.1, to first approximation the contributions from CFFs $\widetilde{\mathcal{H}}$ and $\mathcal{E}$ are negligible in eq.~(\ref{eq:ciunp}) with respect to that of $\mathcal{H}$ since they are kinematically suppressed by an order of magnitude or more. Therefore, in this approximation, the behavior of $\mathcal{C}^{\mathcal{I}}_{\mathrm{unp}}$ is determined by the CFF $\mathcal{H}$, and hence the GPD $H$. Thus, GPD $H$ is constrained by measurements of the beam-helicity asymmetry $\mathcal{A}_{\mathrm{LU}}$.

\section{The {\sc Hermes} experiment in 2006--2007}
\label{sec:experiment}
During the {\sc Hera} winter shutdown 2005/2006, a recoil detector \cite{RecoilTechnicalPaper} was installed in the target region of {\sc Hermes}. The configuration of the forward spectrometer \cite{hermes:spectrometer} remained unchanged. The main purpose of the recoil detector has been the detection of the recoil target proton in order to enhance access to hard exclusive processes at {\sc Hermes}, in particular to DVCS. Data were collected from the recoil detector in conjunction with the forward spectrometer in 2006 and 2007 using the {\sc Hera} electron or positron beam of energy 27.6\,GeV scattering off a target of unpolarized hydrogen or deuterium gas internal to the {\sc Hera} lepton storage ring at {\sc Desy}.

The {\sc Hera} lepton beam was transversely self-polarized by the emission of synchrotron radiation \cite{Sokolov+:1964}. Longitudinal polarization of the beam in the target region was achieved by a pair of spin rotators located upstream and downstream of the experiment \cite{Buon:1986}. The sign of the beam polarization was reversed three times over the running period. Two Compton backscattering polarimeters \cite{TPOL:1994, LPOL:2002} independently measured the longitudinal and transverse beam polarizations.

The commissioning of the recoil detector was completed in 2006 after the switch of the {\sc Hera} lepton beam from electrons to positrons. Therefore, for the analysis of the beam-helicity asymmetry considered here, data collected with only one lepton beam charge but both beam-helicity states are available. For this data set, the average beam polarization was $P_\ell=0.402$ ($-0.394$) for positive (negative) beam helicity, with the total relative uncertainty of 1.96\% \cite{HERApol2012}.

The scattered lepton and particles produced in the polar-angle range $0.04\;\mathrm{rad}<\theta<0.22\;\mathrm{rad}$ were detected by the {\sc Hermes} forward spectrometer. The average lepton-identification efficiency was at least 98\% with hadron contamination of less than 1\%. The produced particles emerging at large polar angles and with small momenta were detected by the {\sc Hermes} recoil detector in the polar-angle range $0.25\;\mathrm{rad}<\theta<1.45\;\mathrm{rad}$, with an azimuthal coverage of about 75\%. The lower-momentum detection threshold for protons was 125\,MeV for this analysis.

\begin{figure}[t]
\centerline{
\epsfig{file=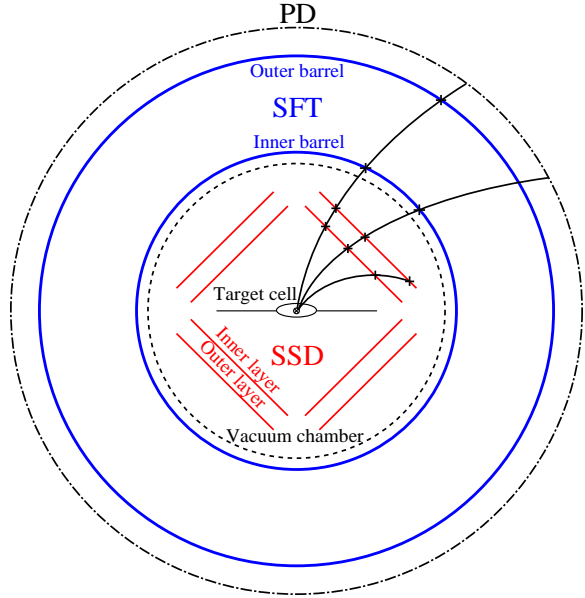, width=0.5\textwidth}
}
\caption{\label{fig:recoildetector} 
Schematic diagram of the {\sc Hermes} recoil detector (cross-section view). The {\sc Hera} lepton beam is perpendicular to the paper plane. The cross section of the target cell is shown as ellipse. The tracking layers are indicated, from inside to outside: inner and outer layers (in diamond shape) of the Silicon Strip Detector (SSD), inner and outer barrels (circles) of the Scintillating Fiber Tracker (SFT). Space-points are indicated by crosses. The SSD modules are located inside the vacuum chamber (dashed circle).
The Photon Detector (PD) shown as a dash-dotted circle is not used in the present analysis. The magnet (not shown) surrounds the detector assembly. Also shown are examples of tracks reconstructed from two, three, and four space-points.
}  
\end{figure}

The recoil detector surrounded the {\sc Hermes} target cell and consisted of several subcomponents embedded in a solenoidal magnetic field with field strength of 1\,T. A detailed description of the recoil-detector components is given in ref.~\cite{RecoilTechnicalPaper}. Figure \ref{fig:recoildetector} shows a schematic view. 
 
The innermost active detector component, surrounding the target cell, was a Silicon Strip Detector (SSD) made up of two layers. Each layer consisted of eight double-sided sensors with orthogonal strips on opposite sides.
In order to minimize the momentum threshold for
proton detection, the amount of passive material between the 75\,$\mu m$ thick target-cell wall and
the first layer of sensors was minimized by placing the SSD and the front-end
read-out electronics inside the {\sc Hera} beam vacuum as close as $5\,\mathrm{cm}$
from the electron or positron beam. Each sensor had an area of $9.9\,\mathrm{cm} \times 9.9\,\mathrm{cm}$
and a strip pitch of $758.2\,\mu\mathrm{m}$ with individual sensor thicknesses
varying between $295\,\mu\mathrm{m}$ and $315\,\mu\mathrm{m}$.

Protons with momenta larger than 250\,MeV passed through  the $1\,\mathrm{mm}$ thick wall of the aluminum vacuum chamber and were detected by the Scintillating Fiber Tracker (SFT). The latter consisted
of two coaxial barrels of scintillating fibers of 1\,mm diameter, the inner barrel at a radius of about
$11.5\,\mathrm{cm}$ and the outer barrel at a radius of about $18.5\,\mathrm{cm}$.
The active length of both barrels was $28\,\mathrm{cm}$. A barrel was
made of two adjacent sub-barrels each consisting of two layers of fibers.
The inner sub-barrel had fibers oriented parallel
to the beam axis, and the outer one had fibers inclined
by $10^\circ$ (stereo layer).

Tracks in the recoil detector are constructed from 3D ``space-points'' in the SSD and the SFT. 
The 1D coordinates from the two sides of a SSD sensor or from adjacent parallel and stereo sub-barrels of a SFT barrel are combined to form 2D coordinates.
For the SSD, this is accomplished taking energy-deposition-correlated combinations of 1D coordinates, while for the SFT all geometrically possible combinations are used. 
The average (sum) of energy deposits for the two 1D coordinates is associated with each 2D coordinate for the SSD (SFT).
Space-points are constructed from 2D coordinates using detector positions. 
Space-point quadruples are obtained as all possible combinations of four space-points, one in each of the two SSD sensors and the two SFT barrels, see figure \ref{fig:recoildetector}. 
A ``geometrical fit'', which only takes into account coordinate information from the space-points, is performed for each such track candidate. This fit uses a helical hypothesis including the lepton-beam axis, and the track candidate is accepted if the $\chi^2$ value
is less than 20. This generous value for four degrees of freedom provides high efficiency
while removing false tracks that arise mainly from space-point quadruples. 
All possible triples or pairs of space-points are similarly treated, whereby space-points belonging to already accepted four-space-point tracks are no longer considered.
The average beam-axis location was determined by fitting 
space-point quadruples, with frequent corrections for beam movement measured independently by
beam-position monitors.

\begin{figure}[t]
 \centerline{
    \epsfig{file=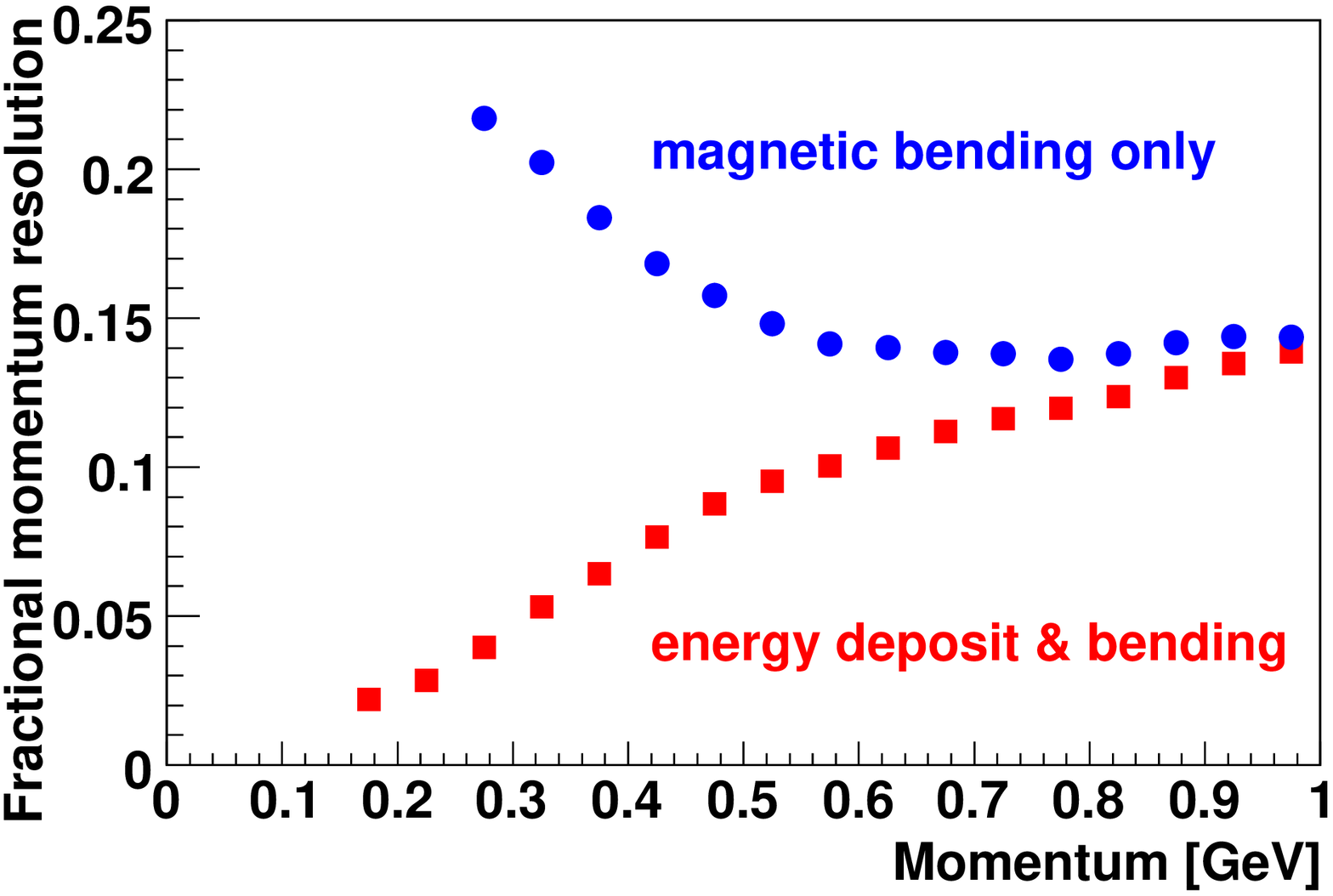,width=0.5\textwidth}
    \epsfig{file=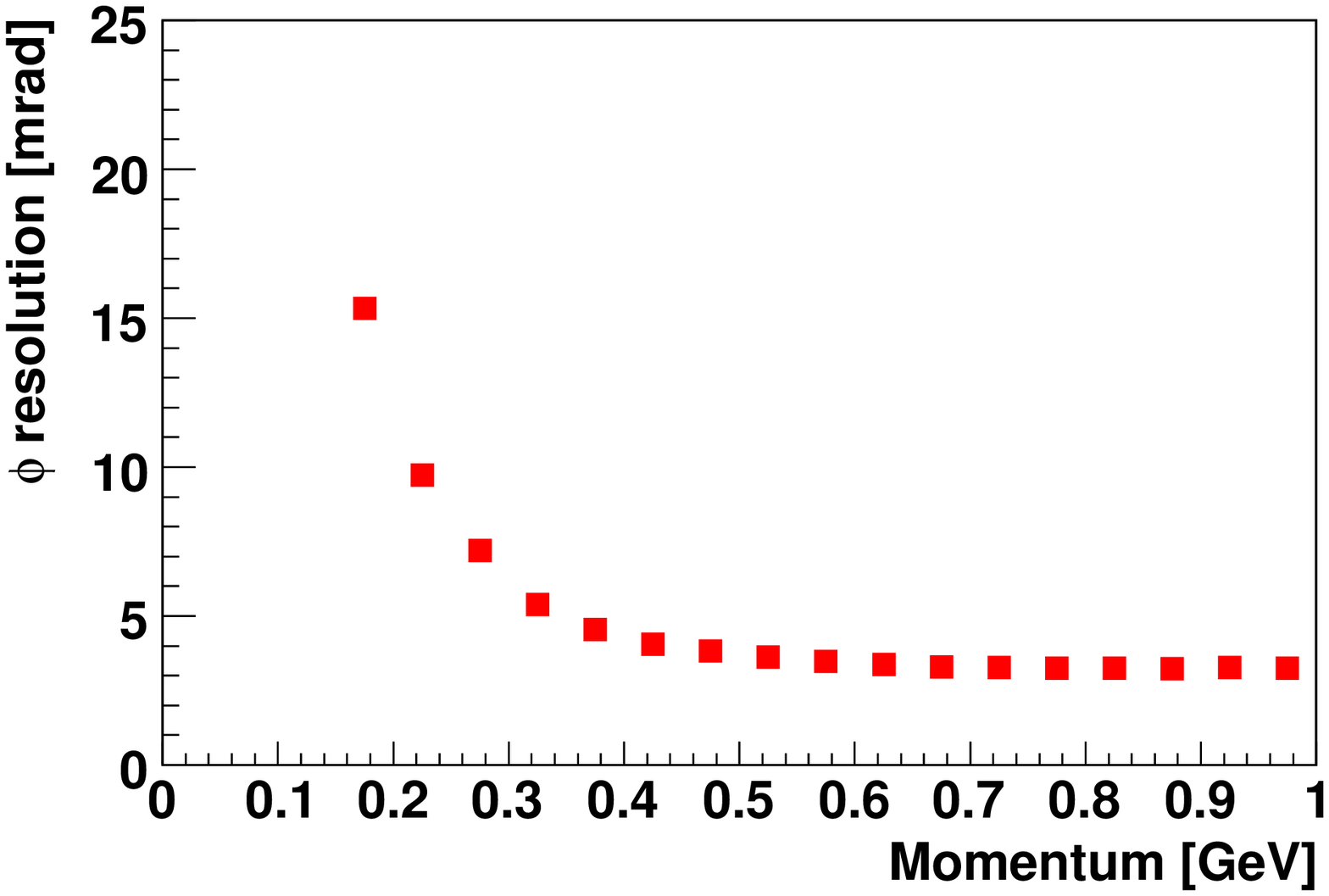,width=0.5\textwidth}
}
\centerline{
    \epsfig{file=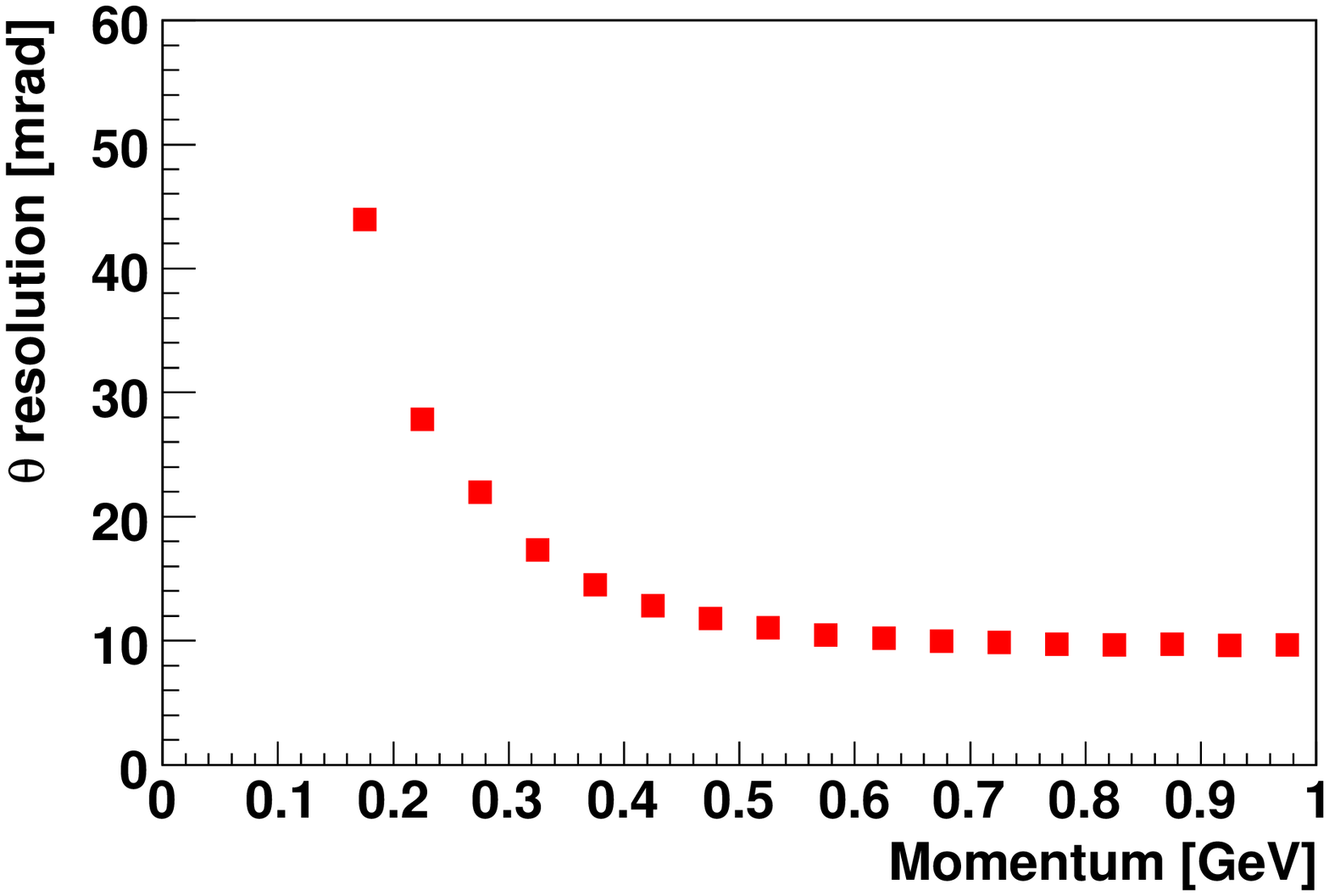,width=0.5\textwidth}
}
    \caption{Top left: momentum resolution versus momentum for proton reconstruction by using only the information on the curvature in the magnetic field
             (circles) and by combining the information on curvature with energy deposition in the SSD (squares). Top right:
	     azimuthal-angle resolution versus momentum. Bottom: polar-angle resolution versus momentum.}
    \label{fig:MomRes}
\end{figure}

The momentum of each track is refitted including energy
deposition in the SSD under the assumption that the particle is a proton, taking into account multiple scattering and energy losses in active and passive material.
If the resulting value of $\chi^2$ exceeds 100 (the optimal value chosen after detailed Monte Carlo studies) the refit is discarded under the assumption that the particle is not a proton. In this case, the track is discarded if there are no space-points in the SFT, otherwise the
momentum reconstruction is based on the geometrical fit.
Momentum-resolution studies were performed based on Monte Carlo data. In figure \ref{fig:MomRes}, the resolution of the momentum and angle
reconstruction is presented for protons.
Reasonable momentum resolution for very low momenta is achieved by combining the information on the curvature in the magnetic field  
with energy depositions in the SSD. The azimuthal- and polar-angle resolution is about 4\,mrad and 10\,mrad, respectively, for proton momenta larger than 0.5\,GeV, deteriorating for lower momenta because of multiple scattering.

Particle identification in the recoil detector, described in detail in ref.~\cite{Xianguo}, is not necessary in this analysis because a clean selection of recoil protons is already accomplished by kinematic event fitting described in the next section.

\section{Event selection}
\label{sec:eventselection}

In this analysis, inclusive $ep\rightarrow eX$ events in the Deep-Inelastic Scattering (DIS) regime are selected by imposing the following kinematic requirements on the identified positron with the largest momentum in the event, as calculated from its four-momentum and that of the incident beam positron: 1\,GeV$^2<Q^2<10$\,GeV$^2$, $W^2>9$\,GeV$^2$, $\nu<22$\,GeV, and $0.03<x_{\mathrm{B}}<0.35$, where $\nu\equiv(pq)/M_p$ is the energy of the virtual photon in the target-rest frame, and $W$ the invariant mass of the $\gamma^*p$ system~\cite{FranksPhD}. 
This sample of inclusive DIS events is employed for determination of relative
luminosities of the two beam-helicity states.

Exclusive $ep\rightarrow ep\gamma$ event candidates are selected from the DIS sample by requiring in the forward spectrometer the detection of exactly one identified positron in the absence of other charged particles and of exactly one signal cluster in the calorimeter not associated with the positron and hence signifying a real photon. The cluster is required to represent an energy deposition above 5\,GeV in the calorimeter and above 1\,MeV in the preshower detector. Two kinematic constraints that were applied in previous {\sc Hermes} DVCS analyses to reduce background are also applied here in order to maintain compatibility and allow direct comparison: i) the polar angle $\theta_{\gamma^*\gamma}$ between the laboratory three-momenta $\vec{q}$ and $\vec{q}^{\;\prime}$ is limited to be less than 45\,mrad, where $\vec{q}$ and $\vec{q}^{\;\prime}$ are the three-momenta of the virtual and real photon, respectively (see figure \ref{fig:phiangle}); ii) the value of $-t$ is limited to be less than 0.7\,GeV$^2$. Here, $-t$ is calculated without use of either the photon-energy measurement or recoil-detector information, under the hypothesis of an exclusive $ep\rightarrow ep\gamma$ event \cite{Air06}:

\begin{equation}
\label{eq:tconstrained}
t=\frac{-Q^2-2\nu(\nu-\sqrt{\nu^2+Q^2}\cos\theta_{\gamma^*\gamma})}{1+ \frac{1}{M_p}(\nu-\sqrt{\nu^2+Q^2}\cos\theta_{\gamma^*\gamma})}.
\end{equation}

\noindent
Moreover, the separation in polar angle between the virtual and real photons is required to be larger than 5\,mrad. This value is determined mainly by the lepton-momentum resolution.

All exclusive event samples considered in this paper are derived from the data set collected in the years 2006/2007 requiring full functionality of the recoil detector.
This data set is a subset of that selected in ref.~\cite{PublicationDraft90} without any such requirement.

A ``pure'' exclusive event sample is selected by combining
information from the recoil detector and forward spectrometer in a kinematic event fit. This fit is based on four-momentum conservation under the hypothesis of the process $ep\rightarrow ep\gamma$. It is performed for every exclusive-event candidate by using the three-momenta of the positron and photon measured in the forward spectrometer and the proton candidate in the recoil detector. The quantity

\begin{equation}
\chi_{\mathrm{kin}}^2 = \sum_{i = 1}^{9}\frac{(r_i^{\mathrm{fit}}-r_i^{\mathrm{meas}})^2}{\sigma_i^2}
\end{equation}

\noindent
is minimized under the four constraints $f_j$ from three-momentum conservation and assumed masses:

\begin{equation}
f_j(r_1^{\mathrm{fit}}, r_2^{\mathrm{fit}}, ..., r_{9}^{\mathrm{fit}}) = 0, \;\;\;j=1,4,
\label{eq:constraints}
\end{equation}

\noindent
where $r_i^{\mathrm{meas}}$ ($r_i^{\mathrm{fit}}$) are measured (fitted) kinematic parameters
of the positron, photon, and the proton candidate and $\sigma_i$ are the measurement uncertainties of
these parameters.
The minimization is conveniently performed using penalty terms:

\begin{equation}
\chi_{\mathrm{pen}}^2 = \sum_{i = 1}^{9}\frac{(r_i^{\mathrm{fit}}-r_i^{\mathrm{meas}})^2}{\sigma_i^2} +
T\cdot \sum_{j=1}^{4}\frac{\left[f_j(r_1^{\mathrm{fit}}, ..., r_{9}^{\mathrm{fit}})\right]^2}{(\sigma^f_{j})^2},
\label{eq:pen}
\end{equation}

\noindent
where $\sigma^f_{j}$ are the propagated uncertainties of $f_j$ and $T$ is a constant number.
For sufficiently large $T$ (the value of $10^8$ is chosen for this analysis), the constraints are automatically satisfied after convergence of 
the minimization procedure. If more than one proton candidate is reconstructed, the one is selected that resulted
in the smallest $\chi_{\mathrm{kin}}^2$ value from the kinematic event fit. The probability calculated from  $\chi_{\mathrm{kin}}^2$ that a particular event satisfied the $ep\rightarrow ep\gamma$ hypothesis is required to be larger than 0.01, a value that is adequate to ensure negligible background contamination. The performance of this event selection is studied using an appropriate mixture of simulated signal and background events \cite{lepto, gmcDVCS} (the simulation is described near the end of this section). Events satisfying all other previously mentioned constraints are found to be selected with high efficiency (83\%) and background contamination less than 0.2\%.
This performance is clearly superior to that from imposing only individual constraints on, 
e.g., the difference between the proton-candidate azimuthal angle or transverse momentum measured by the recoil detector and the expected value of the corresponding variable calculated from the four-momenta of the positron and the real photon detected by the forward spectrometer.

\begin{figure}[t]
\centerline{
\epsfig{file=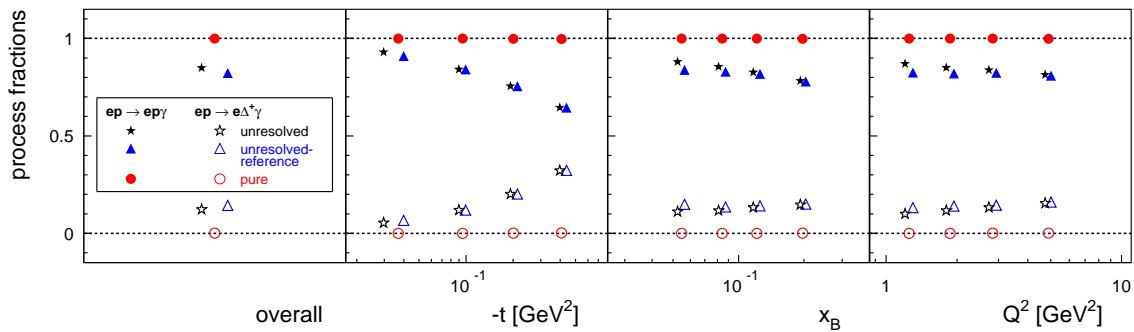, width=\textwidth}
}
\caption{\label{fig:fractions} Fractional contributions from the BH process $ep\rightarrow ep\gamma$ (closed symbols) and the associated BH process $ep\rightarrow e\Delta^+\gamma$ (open symbols), for each of the exclusive samples.  The fractional contributions are extracted from Monte Carlo simulations and are presented in the same kinematic binning as the asymmetry amplitudes in figures \ref{fig:plot-all} and ~\ref{fig:plot-1}. Symbols for the unresolved (unresolved-reference) sample are shifted to the left (right) for better visibility. If the points were plotted without such shifts, a difference would only be visible in the first $-t$ bin.
}
\end{figure}

In the analysis of data collected prior to the installation of the recoil detector, and in the analysis of the present data set without using recoil-detector information, the selection of exclusive $ep\rightarrow ep\gamma$ events is performed by requiring the square of the missing mass
 
\begin{equation}
M_X^2=(k+p-k^\prime-q^\prime)^2 ,
\label{eq:mx2}
\end{equation}

\noindent
calculated using the four-momenta of only the lepton and the real photon,
to be within an ``exclusive region'' about the squared proton mass, with boundaries defined by the resolution of the forward spectrometer: $-(1.5$\,GeV$)^2<M_{X}^2<(1.7$\,GeV$)^2$. Such an event sample includes not only $ep\rightarrow ep\gamma$ events but also contamination from ``associated'' production, $ep\rightarrow eN\pi\gamma$, including resonant production $ep\rightarrow e\Delta^+\gamma$.
This contamination is regarded as unresolved background that remains part of the signal in {\sc Hermes} DVCS analyses that do not use recoil-detector information. (A correction is applied for other background, as described in section~\ref{sec:background}.)
It is estimated using the mixture of simulated events to be about 12\% on average within the exclusive region, as illustrated in figure \ref{fig:fractions}.
Such an exclusive event sample selected by imposing constraints only on the lepton and photon four-momenta is named ``unresolved'' in the following.

In contrast, the analysis of the pure sample, which includes the reconstruction of the recoil proton and kinematic event fitting, introduces two entangled modifications -- a background-free measurement and the kinematic restriction imposed by the acceptance of the recoil detector. 
In order to separate these two effects, the results from the pure sample are compared to results from a subset of the unresolved sample that is subject to the same kinematic restriction.
This ``unresolved-reference'' event sample is selected from the unresolved sample by requiring the missing four-momentum (``hypothetical proton'') to be within the acceptance of the recoil detector.
This requirement results in a loss of about 24\% of the events.
One source of the loss is the effect of the gaps between the SSD modules. The other main source is loss of recoil protons with $p<125$\,MeV, i.e., protons that have too low a momentum to reach the outer layer of the SSD because they are stopped in either the target cell or in the inner layer of the SSD. This lower momentum threshold corresponds to loss of events at low values of $-t<0.016$\,GeV$^2$.
Requiring the proton to be in the recoil-detector acceptance leads to a small modification of the average values $\langle -t\rangle$, $\langle Q^2\rangle$, and $\langle x_{\mathrm{B}}\rangle$  in each kinematic bin compared to the values without such a requirement, as shown in table \ref{tab:averagekine}. As expected by construction of the unresolved-reference sample, the table demonstrates that the average kinematic values of this sample are very similar to those of the pure sample, ensuring that the observables for exclusive photon production are the same for the two samples.

\begin{table}[t]
\begin{center}
\begin{tabular}{c|c||c|c|c||c|c|c||c|c|c}
& & \multicolumn{3}{c||}{unresolved} & \multicolumn{3}{c||}{unresolved-} & \multicolumn{3}{c}{pure} \\
bin & & \multicolumn{3}{c||}{} & \multicolumn{3}{c||}{reference} & \multicolumn{3}{c}{} \\\hline\hline
& $-t$ & $\langle -t\rangle$ & $\langle x_B\rangle$ & $\langle Q^2\rangle$ & $\langle -t\rangle$ & $\langle x_B\rangle$ & $\langle Q^2\rangle$ & $\langle -t\rangle$ & $\langle x_B\rangle$ & $\langle Q^2\rangle$ \\\hline\hline
overall & 0.00--0.70 & 0.116 & 0.097 & 2.53 & 0.136 & 0.101 & 2.63 & 0.129 & 0.102 & 2.63 \\
\hline
1 & 0.00--0.06 & 0.031 & 0.079 & 2.00 & 0.038 & 0.085 & 2.16 & 0.037 & 0.084 & 2.13 \\
2 & 0.06--0.14 & 0.094 & 0.102 & 2.58 & 0.095 & 0.099 & 2.50 & 0.095 & 0.100 & 2.51 \\
3 & 0.14--0.30 & 0.202 & 0.117 & 3.05 & 0.202 & 0.114 & 2.99 & 0.201 & 0.115 & 3.01 \\
4 & 0.30--0.70 & 0.417 & 0.127 & 3.77 & 0.417 & 0.124 & 3.69 & 0.408 & 0.131 & 3.87 \\
\hline\hline
& $x_{\mathrm{B}}$ & $\langle -t\rangle$ & $\langle x_B\rangle$ & $\langle Q^2\rangle$ & $\langle -t\rangle$ & $\langle x_B\rangle$ & $\langle Q^2\rangle$ & $\langle -t\rangle$ & $\langle x_B\rangle$ & $\langle Q^2\rangle$  \\\hline\hline
overall & 0.03--0.35 & 0.116 & 0.097 & 2.53 & 0.136 & 0.101 & 2.63 & 0.129 & 0.102 & 2.63 \\
\hline
1 & 0.03--0.07 & 0.091 & 0.054 & 1.45 & 0.122 & 0.055 & 1.48 & 0.112 & 0.055 & 1.47 \\
2 & 0.07--0.10 & 0.102 & 0.084 & 2.17 & 0.116 & 0.084 & 2.19 & 0.110 & 0.084 & 2.16 \\
3 & 0.10--0.15 & 0.127 & 0.121 & 3.13 & 0.134 & 0.121 & 3.15 & 0.132 & 0.122 & 3.14 \\
4 & 0.15--0.35 & 0.195 & 0.200 & 5.13 & 0.205 & 0.197 & 5.06 & 0.198 & 0.197 & 5.06 \\
\hline\hline
& $Q^2$ & $\langle -t\rangle$ & $\langle x_B\rangle$ & $\langle Q^2\rangle$ & $\langle -t\rangle$ & $\langle x_B\rangle$ & $\langle Q^2\rangle$ & $\langle -t\rangle$ & $\langle x_B\rangle$ & $\langle Q^2\rangle$ \\\hline\hline
overall & 1.00--10.00 & 0.116 & 0.097 & 2.53 & 0.136 & 0.101 & 2.63 & 0.129 & 0.102 & 2.63 \\
\hline
1 & 1.00--1.50 & 0.076 & 0.056 & 1.25 & 0.102 & 0.057 & 1.25 & 0.097 & 0.058 & 1.25 \\
2 & 1.50--2.30 & 0.097 & 0.078 & 1.86 & 0.115 & 0.080 & 1.87 & 0.110 & 0.080 & 1.87 \\
3 & 2.30--3.50 & 0.127 & 0.107 & 2.83 & 0.138 & 0.107 & 2.84 & 0.131 & 0.108 & 2.84 \\
4 & 3.50--10.00 & 0.186 & 0.171 & 4.91 & 0.195 & 0.167 & 4.85 & 0.188 & 0.170 & 4.89 \\
\hline
\end{tabular}
\caption{\label{tab:averagekine}
Average kinematic values for each bin in which the Fourier amplitudes of the beam-helicity asymmetry are extracted, for each of the three exclusive samples. The ``overall'' bin represents a single kinematic bin covering the entire kinematic acceptance of the {\sc Hermes} apparatus. The quantities $-t$ and $Q^2$ are given in units of GeV$^2$.
}
\end{center}
\end{table}

Table \ref{tab:data} summarizes the number of collected events for each of the three exclusive samples: unresolved, unresolved-reference, pure, and the average values of the lepton-beam polarization $P_\ell$. 
The yield of pure events represents about 65\% of the unresolved-reference yield.
Of the total 35\% loss, according to the Monte Carlo studies, the event selection based on kinematic event fitting eliminates from the unresolved-reference sample about 17\% of background events. This also removes 17\% of $ep\rightarrow ep\gamma$ events.
The remaining 1-2\% is attributed to recoil-detector inefficiencies \cite{RecoilTechnicalPaper}.

\begin{table}[t]
\begin{center}
\begin{tabular}{c||c|c|c}
  & $P_\ell>0$ & $P_\ell<0$ & total \\\hline
integrated luminosity  & \ 430\,pb$^{-1}$ \  & \ 240\,pb$^{-1}$ \ & \ 670\,pb$^{-1}$ \ \\\hline
    DIS events ($/10^6$) & 15.8 & 8.7 & 24.5 \\ \hline
unresolved  & 23000 & 12300 & 35300 \\
unresolved-reference  & 17000  & 9200 & 26200 \\
pure  & 11000 & 6000 & 17000 \\\hline
$\langle P_\ell\rangle$ & 0.402 & -0.394 & \ $\langle|P_\ell|\rangle$=0.399 \\
\end{tabular}
\caption{{\sc Hermes} data sets on single-photon production collected with fully commissioned recoil detector in the years 2006 and 2007, using an unpolarized hydrogen target and a positron beam with longitudinal beam polarization $P_\ell$. For each beam helicity separately and for the total data set, the following quantities are given: the integrated luminosity with a systematic uncertainty of 16\% (not used in this analysis), the respective numbers of selected events in the DIS sample used for normalization
and in the three exclusive samples, and the average values of the beam polarization. This polarization has a total relative uncertainty of 1.96\% (dominated by the systematic uncertainty).}
\label{tab:data}
\end{center}
\end{table}

\begin{figure}[t]
\centerline{
\epsfig{file=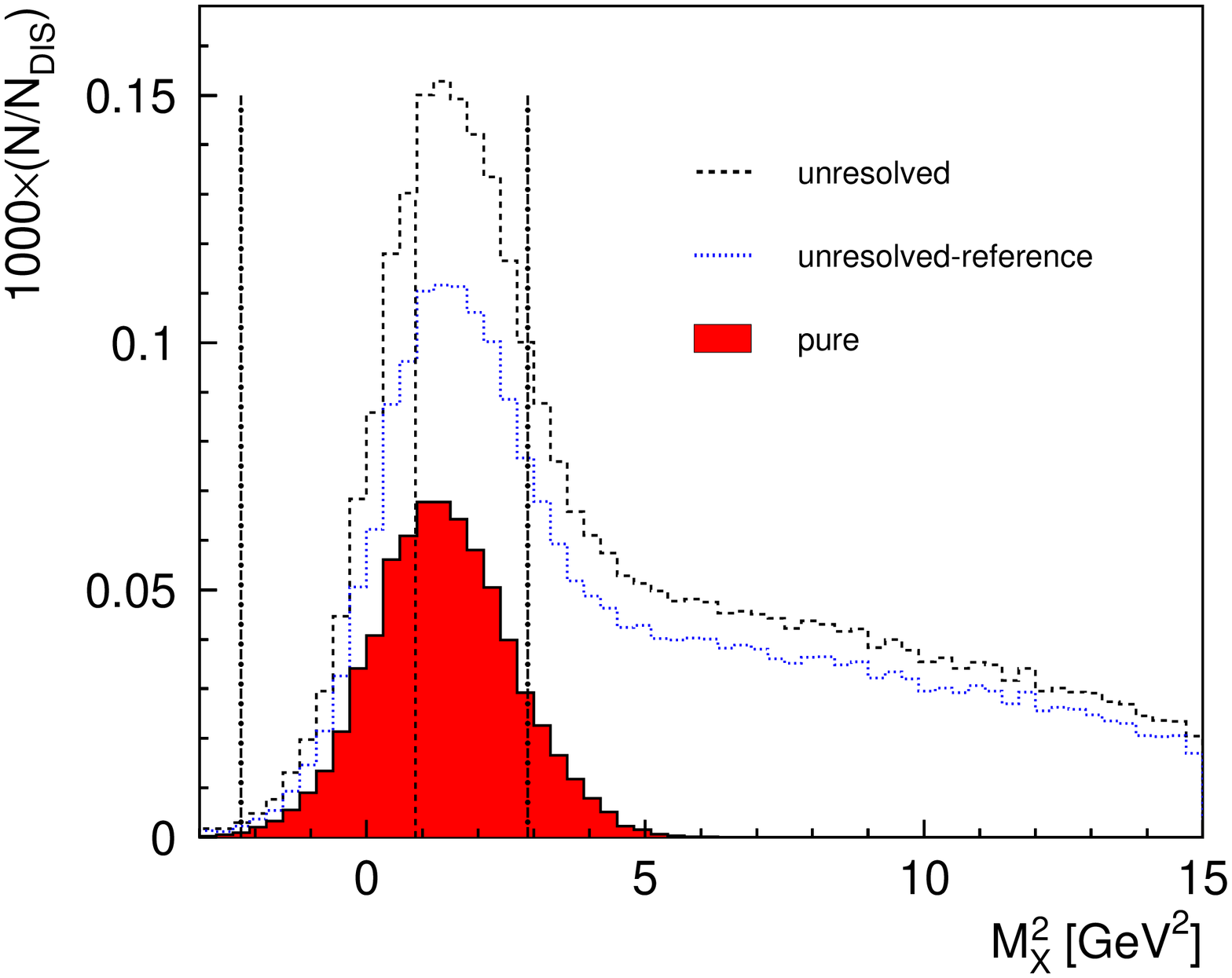,width=0.50\textwidth}
}
\centerline{
\epsfig{file=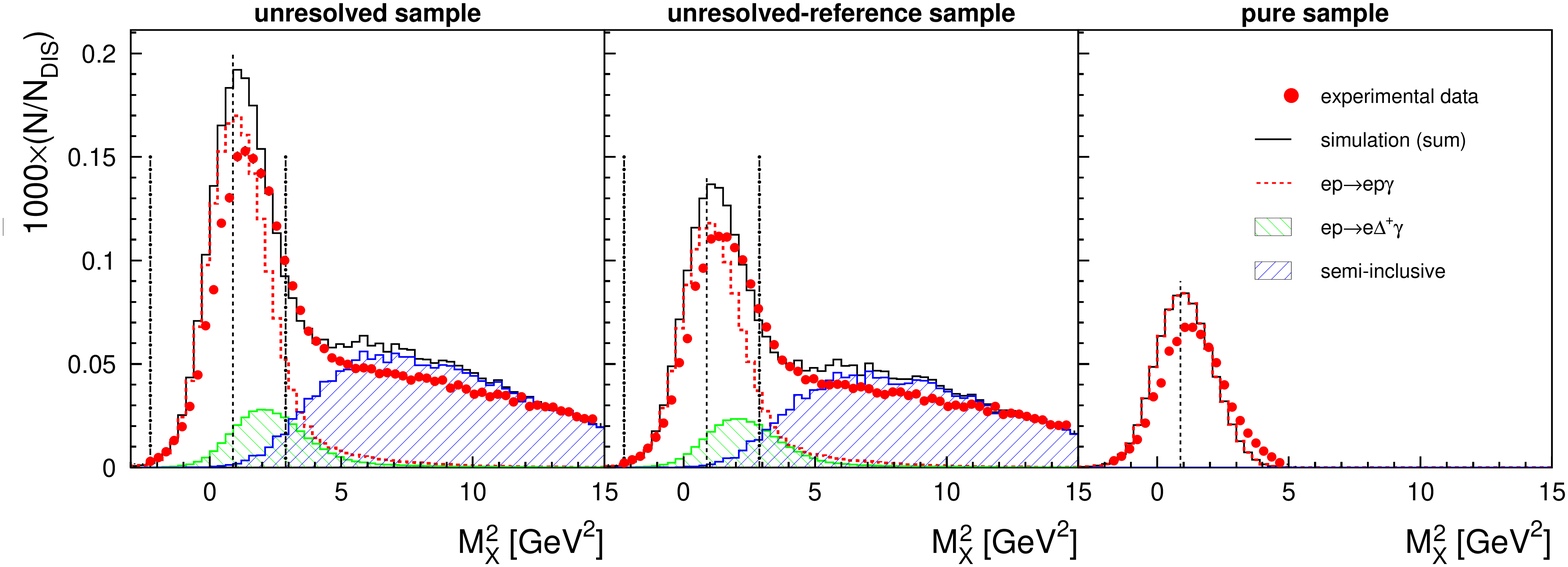,width=\textwidth}
}
\caption{
Distributions of the squared missing mass. The histograms are normalized to the number of DIS events.
Top figure: distributions from experimental data for the three exclusive event samples discussed in the text. 
The requirements applied on the squared missing mass in order to select (only) the unresolved and the unresolved-reference samples are indicated as vertical dashed-dotted lines. The exclusive signal is expected around the square of the proton mass, indicated as vertical dashed line.
Bottom figure, left: unresolved; middle: unresolved-reference; right: pure sample. Experimental data, shown as data points (uncertainties covered by symbols), are compared to simulated data. In every panel, the contribution from BH $ep\rightarrow ep\gamma$ events is indicated as dashed histogram, and the contributions from associated production and semi-inclusive background are shown as hatched histograms. The sum of the simulated distributions is shown as solid histogram. See text for discussion.
}
\label{fig:mx2data}
\end{figure}

Figure \ref{fig:mx2data} shows luminosity-normalized distributions in $M_X^2$ (eq.~(\ref{eq:mx2})) for each of the three exclusive samples. The figure also presents a comparison of experimental data to a mixture of simulated data samples. 
Bethe--Heitler events are simulated  using the Mo--Tsai formalism \cite{motsai}, by an event generator based on ref.~\cite{gmcDVCS} and described in detail in ref.~\cite{BernhardsPhD}. This sample of BH events includes events from associated production generated using the parameterization of the form factor for the resonance region from ref.~\cite{Bra76}. (The DVCS process is not included since an event generator for associated production in DVCS is unavailable.)
Semi-inclusive events are simulated by using an event generator based on LEPTO \cite{lepto}, including the RADGEN \cite{radgen} package for radiative effects.
The Monte-Carlo yield exceeds the experimental data by about 20\% in the exclusive region, as observed in previous studies of {\sc Hermes} data \cite{PublicationDraft68}.
The rightmost panel of figure \ref{fig:mx2data} demonstrates that the simulation describes the pure sample well enough to validate the negligible estimate of background in this sample.

The fractional contributions of the reaction $ep\rightarrow ep\gamma$ and associated processes, determined from the aforementioned  mixture of simulated signal and background events,
are detailed in figure \ref{fig:fractions} in each kinematic bin in which the asymmetry amplitudes are extracted. For the pure sample, the contribution of the process $ep\rightarrow ep\gamma$ is found to be close to 100\% and the contribution of events from associated processes is close to zero in all kinematic bins. 
In contrast, for the unresolved and unresolved-reference sample the contribution of the associated process is on average about 12\% and 14\%, respectively, rising with increasing values of $-t$;
i.e., imposing the acceptance of the recoil detector on the unresolved sample has little effect on the background fractions.
Therefore, comparison of the results from the pure and unresolved-reference samples demonstrates the effects of elimination of associated background without changing the experimental acceptance.

\section{Extraction of single-charge beam-helicity asymmetry amplitudes}
\label{sec:formalism}
Fourier amplitudes of the single-charge beam-helicity asymmetry $\mathcal{A}_{\mathrm{LU}}(\phi;e_\ell)$ of eq.~(\ref{eq:alu_sbc1}) are extracted from each of the samples described in section~\ref{sec:eventselection}. The extraction formalism is described in more detail in ref.~\cite{PublicationDraft68}. It is based on a maximum-likelihood technique \cite{MML}, which provides a bin-free fit in the azimuthal angle $\phi$. Event weights are employed in the fit in order to account for luminosity imbalances with respect to beam polarization.

Based on eq.~(\ref{eq:sigma_lu}), the distribution of the expectation value of the yield for scattering of a longitudinally polarized positron beam from an unpolarized hydrogen target is given by

\begin{equation}
\langle\mathcal{N}\rangle(\phi; e_\ell, P_\ell)=\mathcal{L}(e_\ell, P_\ell)\eta(\phi)\sigma_{\mathrm{UU}}(\phi)\left[1+P_\ell\mathcal{A}_{\mathrm{LU}}(\phi; e_\ell)\right],
\label{eq:yield}
\end{equation}

\noindent where $\mathcal{L}$ denotes the integrated luminosity determined by counting inclusive DIS events and $\eta$ the detection efficiency. Here it is assumed that the polarization-dependent cross sections depend linearly on the kinematic variables over which the yield is integrated. (A systematic uncertainty associated with this assumption is discussed in the next section.) The asymmetry $\mathcal{A}_{\mathrm{LU}}(\phi;e_\ell)$ is expanded in terms of harmonics in $\phi$ in order to extract azimuthal asymmetry amplitudes: 

\begin{equation}
\mathcal{A}_{\mathrm{LU}}(\phi;e_\ell)\simeq A_{\mathrm{LU}}^{\sin\phi}\sin\phi+A_{\mathrm{LU}}^{\sin(2\phi)}\sin(2\phi),
\label{eq:aexpansion}
\end{equation}

\noindent 
where the approximation is due to the truncation of the infinite Fourier series.
Note that $A_{\mathrm{LU}}^{\sin\phi}$ is related, but not identical to $s_1^{\mathcal{I}}$ since there is an additional $\phi$-dependence in the lepton propagators in eq.~(\ref{eq:alu_sbc2}), and there is another $\sin\phi$ amplitude $s_1^{\mathrm{DVCS}}$ in eq.~(\ref{eq:alu_sbc2}). The former statement also holds for $A_{\mathrm{LU}}^{\sin(2\phi)}$ and $s_2^{\mathcal{I}}$.

As a consistency check for extraneous harmonics caused by the lepton propagators in eq.~(\ref{eq:alu_sbc2}) and as a test of the normalization of the fit, the maximum likelihood fit was repeated including the terms $A_{\mathrm{LU}}^{\cos(0\phi)}$ and $A_{\mathrm{LU}}^{\cos\phi}$. As expected, these spurious terms were found to be compatible with zero within statistical uncertainties and have negligible impact on the resulting asymmetry amplitudes.
This provides evidence that the experimental acceptance did not suffer instabilities correlated with beam helicity.

\section{Background corrections and systematic uncertainties}
\label{sec:background}

\subsection{Corrections and uncertainty contributions for the unresolved samples}
The asymmetry amplitudes extracted from the unresolved and unresolved-reference samples are corrected for the presence of background that involves semi-inclusive or hard-exclusive neutral pseudo-scalar meson production (mainly $\pi^0$), where one of the two photons from the meson decay escapes the acceptance of the calorimeter, or the two photons are registered as one calorimeter cluster, thus faking a single-photon event candidate.
In order to correct the measured amplitude $A_{\textrm{meas}}$ for these background processes, the following procedure is applied in every kinematic bin to obtain the asymmetry amplitude $A_{\textrm{final}}$ corrected for background:

\begin{equation}
\label{eq:bgcorr}
A_{\textrm{final}} = \frac{A_{\textrm{meas}} - f_{\textrm{semi}}A_{\textrm{semi}} - f_{\textrm{excl}}A_{\textrm{excl}}}{1 - f_{\textrm{semi}} - f_{\textrm{excl}}},
\end{equation}

\noindent where $f_{\textrm{semi}}$ $(f_{\textrm{excl}})$ is the fraction of semi-inclusive (exclusive)
$\pi^{0}$ events in the data sample and $A_{\textrm{semi}}$ $(A_{\textrm{excl}})$
is the corresponding asymmetry amplitude.
These background fractions are determined from Monte Carlo simulations.

For the estimate of the semi-inclusive background fraction, the event generator LEPTO \cite{lepto} is employed, yielding about 2.7\% for the unresolved sample and about 3.1\% for the unresolved-reference sample, with only weak kinematic dependence.
The background asymmetry amplitude $A_{\textrm{semi}}$ is extracted from experimental data. Neutral pions are reconstructed from a sample of events where two photons are detected by requiring the invariant mass of the two photons to be close to the mass of the neutral pion: 0.1\,GeV$<M_{\gamma \gamma}< 0.17$\,GeV. In addition, the fractional energy carried by the neutral pions is required to be large, $E_{\pi^0}/\nu>0.8$, as only these contribute
to the exclusive region according to simulations \cite{PhDZhenyu}. Simulations showed that
the extracted $\pi^0$ asymmetry does not depend on whether only one or both photons are
in the acceptance. 
Asymmetry amplitudes are extracted from the resulting two-photon data sample using the same fit function
as that used to extract the asymmetry amplitudes for the exclusive $ep\rightarrow ep\gamma$ measurements.

Hard-exclusive neutral pseudo-scalar meson production was found to be undetectable at {\sc Hermes} without using the recoil detector \cite{PhDArne}, and its isolation is found to be difficult even with its use. Therefore, the asymmetry is assumed to be equally probable over the range from -1 to 1 and a value of $0\pm\frac{2}{\sqrt{12}}$ is assigned to the background asymmetry amplitude $A_{\textrm{excl}}$. For the estimate of the fraction of exclusive $\pi^0$ events, an exclusive event generator is used that is described in greater detail in ref.~\cite{exclpion}.
For both unresolved samples, the contribution from hard-exclusive meson production $f_{\mathrm{excl}}$ is about 0.4\%.

The systematic uncertainty for a given corrected amplitude due to the background (bg) contamination is taken as one half of the correction to the amplitude: $\delta A_{\mathrm{syst.}}^{\mathrm{bg}}=\frac{1}{2}|A_{\mathrm{final}}-A_{\mathrm{meas}}|$. This is considered to be sufficient to account for the mismatch of measured and simulated shape (and size) of the semi-inclusive background observed in figure \ref{fig:mx2data}.
The statistical uncertainty on the total correction due to the estimated background asymmetry amplitudes $A_{\textrm{semi}}$ and $A_{\textrm{excl}}$ is propagated as a contribution to the final statistical uncertainty.

After applying eq.~(\ref{eq:bgcorr}), the asymmetry amplitude $A_{\mathrm{final}}$ extracted from the unresolved or unresolved-reference event samples is expected to originate only from $ep\rightarrow ep\gamma$ and associated processes. No correction is applied for the contamination by these associated processes because the relevant asymmetry amplitudes are not known, i.e., for these samples the contribution of associated production is considered to be part of the signal (see also figure \ref{fig:fractions}).

\subsection{Corrections and uncertainty contributions for the pure sample}

The asymmetry amplitudes extracted from the pure event sample are not corrected for background because the estimated contribution to the yield from background processes is negligible: on average about 0.015\% from semi-inclusive processes and less than $0.2\%$ from associated processes.

\subsection{Uncertainty contributions common to all samples}
Systematic uncertainties arising from the forward-spectrometer acceptance, smearing, and finite bin width have been studied for several previous DVCS analyses (see, e.g., ref.~\cite{PublicationDraft68}). 
An ``all-in-one'' estimate of the combination of these uncertainties was designed to account for possible discrepancies between an evaluation of model asymmetries at the measured mean kinematic values and an evaluation accounting for the experimental limitations listed above using a full simulation of the detector.
A similar approach is taken here, except that only variant A of ref.~\cite{gmcDVCS} is employed, which is based on the model of ref.~\cite{Vanderhaeghen:1999xj}. This variant describes the asymmetry amplitudes extracted from the pure sample well, while the other variants employed in ref.~\cite{PublicationDraft68} do not. Events from associated production are not included for this study.
Even though the results are plotted versus mean values of measured kinematic quantities, the experimental acceptance can influence the results because of a non-linear dependence of the observables on those kinematic quantities.
The effect of the acceptance of the recoil detector on the results from the pure sample is included in this study. 
Because of an improved survey of the apparatus, effects of a misalignment of the forward spectrometer are considered to be negligible.
Studies have shown that possible misalignments of the recoil detector affect only the efficiency of the constraint from the kinematic event fit and not the resulting asymmetry amplitudes. 
The effects of inefficiencies in the trigger scintillator hodoscopes of the forward spectrometer and in the recoil detector are found to be negligible. Asymmetry results for the pure sample obtained using one of the four quadrants of the recoil detector at a time are found to be consistent within statistical uncertainties.
No systematic uncertainty was assigned for the relative luminosity measurement based on counting inclusive DIS events because possible changes in spectrometer-detection efficiencies correlated with beam helicity would tend to cancel in the extracted asymmetries, and because the extracted $\cos(0\phi)$ asymmetry amplitude was found to be consistent with zero, as mentioned at the end of section~\ref{sec:formalism}.

\subsection{Summary of systematic uncertainties}

The individual contributions to the systematic uncertainty of each of the asymmetry amplitudes are added in quadrature to obtain the final systematic uncertainty in every bin in which the amplitudes are extracted. This procedure is applied separately for each of the three exclusive event samples. Table \ref{tab:sys} gives the results of the background-corrected asymmetry amplitudes extracted in a single kinematic bin covering the entire kinematic acceptance of the {\sc Hermes} apparatus, together with their statistical and systematic uncertainties and the individual contributions to the latter.
The dominant contribution to the systematic uncertainty is the ``all-in-one'' uncertainty. 
(This contribution is larger than that estimated in refs.~\cite{PublicationDraft69, PublicationDraft90} because of the use of a different theoretical model.)

\begin{table}[t]
\begin{center}
\begin{tabular}[h]{c|c||c|c|c||c|c}
amplitude & data sample & $A$ & $\delta A_{\mathrm{stat.}}$ & $\delta A_{\mathrm{syst.}}$ &  $\delta A_{\mathrm{syst.}}^{\mathrm{all-in-one}}$ & $\delta A_{\mathrm{syst.}}^{\mathrm{bg}}$  \\ \hline
& unresolved & -0.250 & 0.019 & 0.047 & 0.047 & 0.004 \\
$A_{\mathrm{LU}}^{\sin\phi}$ & unresolved-reference & -0.274 & 0.022 & 0.037 & 0.036 & 0.005 \\
& pure  & -0.328  & 0.027 & 0.045 & 0.045 & -\\
\hline
& unresolved & 0.004 & 0.019 & 0.004 & 0.004 & $<$0.001 \\
$A_{\mathrm{LU}}^{\sin(2\phi)}$ & unresolved-reference & 0.011 & 0.021 & 0.004 & 0.004 & $<$0.001 \\
& pure  & 0.014  & 0.026 & 0.002 & 0.002 & -\\
\end{tabular}
\caption{
Asymmetry amplitudes extracted from each of the three exclusive event samples in the ``overall'' kinematic bin together with their statistical and systematic uncertainties. Also the individual contributions to the latter are given, ``bg'' indicating the uncertainty arising from the background correction, and ``all-in-one'' the combined uncertainty arising from detector acceptance, smearing, and finite bin width. The amplitudes are corrected for background from semi-inclusive and exclusive neutral pions where applicable. A separate scale uncertainty arising from the measurement of the beam polarization amounts to 1.96\%.
}
\label{tab:sys}
\end{center}
\end{table}

There is a separate scale uncertainty of 1.96\% arising from the uncertainty
in the measurement of the beam polarization. This uncertainty is not included in the error bands used to display the combined systematic uncertainty.

\section{Results and discussion}
\label{sec:results}

Figure \ref{fig:plot-all} and table \ref{tab:results} show the $\sin(n\phi)$ amplitudes of the single-charge beam-helicity asymmetry extracted from 2006 and 2007 hydrogen data collected with a positron beam and recoil detector. The results are displayed in projections versus $-t$, $x_{\textrm{B}}$, and $Q^2$ and also in a single kinematic bin covering the entire kinematic acceptance of the {\sc Hermes} apparatus (``overall''). For the extraction and presentation of asymmetry amplitudes, kinematic variables measured by only the forward spectrometer are used. The calculation of $x_{\mathrm{B}}$ and $Q^2$ requires the identification and momentum measurement of the scattered lepton. Also for the calculation of $-t$ (see eq.~(\ref{eq:tconstrained})), only the measured kinematic parameters of the lepton are used.

\subsection{Results for the pure sample}
\label{sec:resultspure}

\begin{figure}[t]
\centerline{
\epsfig{file=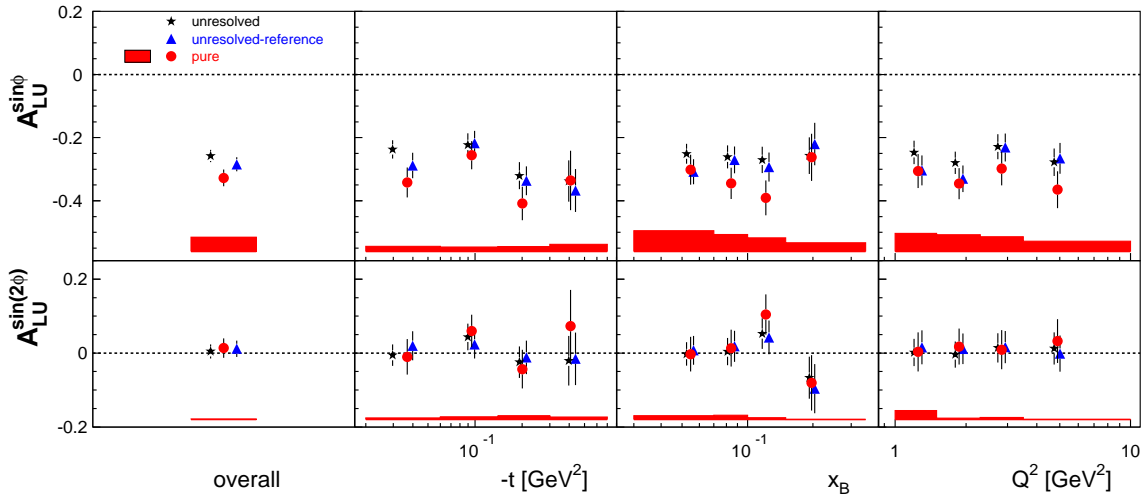, width=\textwidth}
}
\caption{\label{fig:plot-all} 
 Amplitudes of the single-charge beam-helicity asymmetry in deeply virtual Compton scattering shown in projections of $-t$, $x_{\textrm{B}}$, and $Q^2$. The ``overall'' results shown in the very left panel are extracted in a single kinematic bin covering the entire kinematic acceptance.
Statistical uncertainties are shown by error bars. The bands represent the systematic uncertainties of the amplitudes extracted from the pure sample.
A separate scale uncertainty arising from the measurement of the beam polarization amounts to 1.96\%.
Shown are amplitudes extracted from 
a) the pure $ep\rightarrow ep\gamma$ sample (red circles, shown at their kinematic values), i.e., obtained with recoil-proton reconstruction; 
b) the unresolved-reference sample (blue triangles, shifted to the right for better visibility), i.e., without recoil-proton reconstruction but requiring its four-momentum to be in the recoil-detector acceptance; 
c) the unresolved sample (black stars, shifted to the left for better visibility), i.e., without requirements from recoil-detector acceptance and reconstruction. The actually reconstructed kinematic values are specified in table \ref{tab:averagekine} for every bin in which the amplitudes are presented. 
The latter two sets of amplitudes are subject to an average contribution of 14\% and 12\%, respectively, for associated processes (see figure \ref{fig:fractions} for the kinematic dependences). All three sets of amplitudes are extracted from the same 2006/2007 positron-beam data set and the results are strongly statistically correlated. 
}
\end{figure}

In figure \ref{fig:plot-all}, the results indicated by the circles are extracted from the pure $ep\rightarrow ep\gamma$ event sample
defined in section~\ref{sec:eventselection}, i.e., they involve the reconstruction of the recoiling proton and kinematic event fitting. The other sets of data points will be discussed below.
The overall value of the leading $\sin\phi$ amplitude is negative and significantly different from zero. Its one-dimensional projections in $x_{\textrm{B}}$ and $Q^2$ (which are highly correlated) reveal no dependences. 
There is no clear indication for a dependence on $-t$, although this amplitude is expected to approach zero as $-t$ approaches zero (see eq.~(\ref{eq:s1ciunp})).
The overall value of the sub-leading $\sin(2\phi)$ amplitude is compatible with zero within its total experimental uncertainty.

\subsection{Comparison to results without recoil-proton detection}
\label{sec:resultsunresolved}

In order to demonstrate the effects of contributions by associated processes that were included in previous {\sc Hermes} measurements of the beam-helicity asymmetry, figure \ref{fig:plot-all} also shows the data points for the $\sin(n\phi)$ amplitudes extracted from the unresolved and unresolved-reference samples.
The comparison is performed in two steps in order to isolate these effects from those of the change in the experimental acceptance due to the recoil detector.

i) The triangles represent data points extracted from the unresolved-reference sample, which is obtained by a missing-mass analysis without using recoil-detector information, but requiring the hypothetical recoil proton to be in the acceptance of the recoil detector. The comparison with the amplitudes extracted from the pure sample demonstrates the change of the measured amplitudes arising from only the removal of the events from associated production, because both sets of amplitudes are measured within the same acceptance of forward spectrometer combined with recoil detector.
There is an indication that the overall value of the $\sin\phi$ amplitude for the pure sample, $-0.328\pm0.027\;\mathrm{(stat.)}$, is larger in magnitude than that of the unresolved-reference sample by $0.054\pm0.016\;\mathrm{(stat.)}$. The calculation of the latter uncertainty is based on the good approximation that one sample is a sub-sample of the other.
The systematic uncertainties are not relevant for this
comparison because the dominant contributions are fully correlated.

ii)  The stars represent data points extracted from the unresolved sample, which is obtained by a missing-mass analysis without using recoil-detector information. Comparison with the amplitudes extracted from the unresolved-reference sample from i) demonstrates the impact of the recoil-detector acceptance on the observed amplitudes. 
These two sets of amplitudes are subject to very similar background conditions as discussed near the end of section \ref{sec:eventselection}.
The overall value of the $\sin\phi$ amplitude extracted from the unresolved-reference sample, $-0.274\pm0.022\;\mathrm{(stat.)}$,
is observed to be slightly larger in magnitude in comparison with that extracted 
from the unresolved sample, amounting to a difference of $0.024\pm0.011\;\mathrm{(stat.)}$. The calculation of this uncertainty accounts for fully correlated data samples.

As elaborated in section~\ref{sec:eventselection}, the lower-momentum threshold that arises from imposing the recoil-detector acceptance results in a loss of acceptance at low values of $-t$, which is reflected in the larger statistical uncertainty for the amplitude extracted from the unresolved-reference sample in the lowest $-t$ bin in figure \ref{fig:plot-all}. 

The  overall $A_{\mathrm{LU}}^{\sin\phi}$ result for the unresolved sample, being representative of previous {\sc Hermes} publications, is $-0.250\pm0.019\;\mathrm{(stat.)}$ versus  $-0.328\pm0.027\;\mathrm{(stat.)}$ for the pure sample. 
The results for the overall $\sin\phi$ asymmetry amplitude for the unresolved and pure samples differ by $0.078\pm0.019\;\mathrm{(stat.)}$, arising from both the acceptance of the recoil detector and the elimination of background from associated production.

The sub-leading $\sin(2\phi)$-amplitude is found to be compatible with zero within total experimental uncertainties for all three event samples. 

\subsection{Comparison with theory}
\label{sec:theory}

\begin{figure}[t]
\centerline{
\epsfig{file=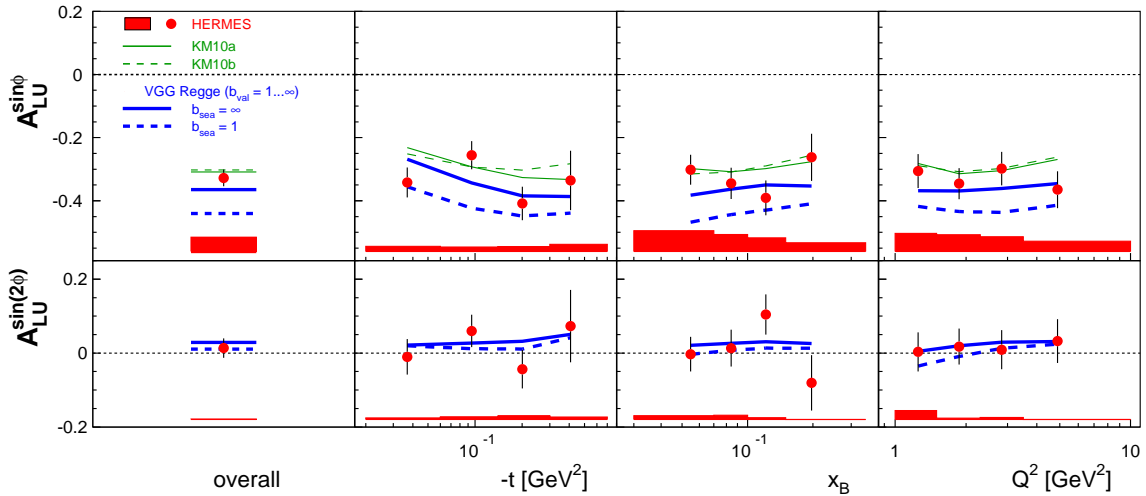, width=\textwidth}
}
\caption{\label{fig:plot-1}  Amplitudes of the single-charge beam-helicity asymmetry extracted from the pure $ep\rightarrow ep\gamma$ sample obtained with recoil-proton reconstruction. 
The amplitudes are presented in projections of $-t$, $x_{\textrm{B}}$, and $Q^2$. The ``overall'' results shown in the very left panel are extracted in a single kinematic bin covering the entire kinematic acceptance.
Statistical (systematic) uncertainties are represented by error bars (bands).
A separate scale uncertainty arising from the measurement of the beam polarization amounts to 1.96\%.
The theoretical models, which are described in the text, are evaluated at the average values of the kinematic parameters specified in table \ref{tab:averagekine} (the points are interpolated by straight lines). The thickness of the VGG lines represents the range $b_{\mathrm{val}}= 1...\infty$.}
\end{figure}

In figure \ref{fig:plot-1}, the asymmetry amplitudes extracted from the pure sample are compared with calculations \cite{Vdhcode}, labeled ``VGG Regge", from the GPD model described in ref.~\cite{Vanderhaeghen:1999xj}. 
Variants of this model differ in the $t$ dependence of GPD $H$. Here, a Regge-inspired ansatz for the $t$ dependence is used.  The skewness dependence is controlled by the $b$ parameter, where  $b_{\mathrm{val}}$ ($b_{\mathrm{sea}}$) is a free parameter for the valence (sea) quarks.
 The result of the model calculation depends only very weakly on the value of $b_{\mathrm{val}}$.
For the sea quarks, the skewness-independent variant of the model ($b_\mathrm{sea}=\infty$)  is consistent with the data, while a maximal skewness dependence ($b_\mathrm{sea}=1$) is disfavored.

Also shown in the figure are the results from model calculations \cite{Kum09} labeled ``KM''.   This model is a dual representation of GPDs with very weakly entangled skewness and $t$ dependences.  The $t$ dependence is approximated by a physically motivated Regge dependence. The model is constrained by previous measurements at {\sc Hermes}, Jefferson Lab, and the collider experiments at H{\sc era}. The fits resulting in the solid curves disregard data from experiments at Hall A at Jefferson Lab (KM10a), while the dashed curves include these data (KM10b). The KM calculations agree well with the extracted leading amplitude. 

The observed difference between the asymmetries extracted from the pure and the unresolved-reference samples is qualitatively consistent with that predicted by a model calculation \cite{AssoAsy} using a soft pion theorem based on chiral symmetry and a $\Delta$(1232)-resonance model using the large $N_c$ limit to relate the GPDs for $\Delta$ excitation to those for the nucleon ground state. In this comparison, it is important to note that the kinematic conditions for this model correspond approximately to the third $-t$ bin of the present measurement.

\begin{table}[t]
\begin{center}
\begin{tabular}{c||c|c|c}
kinematic bin & unresolved & unresolved-reference & pure   \\\hline\hline
& \multicolumn{3}{|c}{$A_{\mathrm{LU}}^{\sin\phi}$} \\\hline
overall bin & -0.250 $\pm$ 0.019 $\pm$ 0.047 & -0.274 $\pm$ 0.022 $\pm$ 0.037 & -0.328 $\pm$ 0.027 $\pm$ 0.045\\
\hline \hline
0.00 $<-t\le$ 0.06 & -0.234 $\pm$ 0.029 $\pm$ 0.013 & -0.282 $\pm$ 0.039 $\pm$ 0.008 & -0.342 $\pm$ 0.047 $\pm$ 0.016\\
0.06 $<-t\le$ 0.14 & -0.215 $\pm$ 0.035 $\pm$ 0.010 & -0.210 $\pm$ 0.038 $\pm$ 0.011 & -0.256 $\pm$ 0.045 $\pm$ 0.014\\
0.14 $<-t\le$ 0.30 & -0.305 $\pm$ 0.041 $\pm$ 0.015 & -0.321 $\pm$ 0.044 $\pm$ 0.015 & -0.409 $\pm$ 0.053 $\pm$ 0.015\\
0.30 $<-t\le$ 0.70 & -0.324 $\pm$ 0.063 $\pm$ 0.019 & -0.353 $\pm$ 0.066 $\pm$ 0.020 & -0.336 $\pm$ 0.094 $\pm$ 0.022\\
\hline \hline
0.03 $<x_{\mathrm{B}}\le$ 0.07 & -0.249 $\pm$ 0.031 $\pm$ 0.065 & -0.304 $\pm$ 0.039 $\pm$ 0.051 & -0.302 $\pm$ 0.047 $\pm$ 0.066\\
0.07 $<x_{\mathrm{B}}\le$ 0.10 & -0.254 $\pm$ 0.036 $\pm$ 0.050 & -0.262 $\pm$ 0.041 $\pm$ 0.045 & -0.345 $\pm$ 0.050 $\pm$ 0.053\\
0.10 $<x_{\mathrm{B}}\le$ 0.15 & -0.260 $\pm$ 0.040 $\pm$ 0.041 & -0.281 $\pm$ 0.044 $\pm$ 0.038 & -0.391 $\pm$ 0.055 $\pm$ 0.043\\
0.15 $<x_{\mathrm{B}}\le$ 0.35 & -0.234 $\pm$ 0.054 $\pm$ 0.030 & -0.200 $\pm$ 0.062 $\pm$ 0.028 & -0.262 $\pm$ 0.075 $\pm$ 0.027\\
\hline \hline
1.00 $<Q^2\le$ 1.50 & -0.239 $\pm$ 0.036 $\pm$ 0.062 & -0.291 $\pm$ 0.045 $\pm$ 0.044 & -0.306 $\pm$ 0.054 $\pm$ 0.057\\
1.50 $<Q^2\le$ 2.30 & -0.271 $\pm$ 0.034 $\pm$ 0.054 & -0.318 $\pm$ 0.040 $\pm$ 0.044 & -0.346 $\pm$ 0.049 $\pm$ 0.053\\
2.30 $<Q^2\le$ 3.50 & -0.222 $\pm$ 0.038 $\pm$ 0.045 & -0.223 $\pm$ 0.043 $\pm$ 0.042 & -0.298 $\pm$ 0.053 $\pm$ 0.047\\
3.50 $<Q^2\le$ 10.00 & -0.268 $\pm$ 0.042 $\pm$ 0.033 & -0.256 $\pm$ 0.047 $\pm$ 0.031 & -0.365 $\pm$ 0.058 $\pm$ 0.032\\
\hline \hline
& \multicolumn{3}{|c}{$A_{\mathrm{LU}}^{\sin(2\phi)}$} \\\hline
overall bin & 0.004 $\pm$ 0.019 $\pm$ 0.004 & 0.011 $\pm$ 0.021 $\pm$ 0.004 & 0.014 $\pm$ 0.026 $\pm$ 0.002\\
\hline \hline
0.00 $<-t\le$ 0.06 & -0.003 $\pm$ 0.029 $\pm$ 0.006 & 0.023 $\pm$ 0.038 $\pm$ 0.008 & -0.010 $\pm$ 0.048 $\pm$ 0.005\\
0.06 $<-t\le$ 0.14 & 0.038 $\pm$ 0.034 $\pm$ 0.007 & 0.019 $\pm$ 0.037 $\pm$ 0.007 & 0.060 $\pm$ 0.044 $\pm$ 0.008\\
0.14 $<-t\le$ 0.30 & -0.022 $\pm$ 0.041 $\pm$ 0.009 & -0.009 $\pm$ 0.043 $\pm$ 0.009 & -0.044 $\pm$ 0.052 $\pm$ 0.011\\
0.30 $<-t\le$ 0.70 & -0.023 $\pm$ 0.065 $\pm$ 0.002 & -0.018 $\pm$ 0.069 $\pm$ 0.003 & 0.073 $\pm$ 0.098 $\pm$ 0.007\\
\hline \hline
0.03 $<x_{\mathrm{B}}\le$ 0.07 & -0.003 $\pm$ 0.031 $\pm$ 0.005 & 0.007 $\pm$ 0.038 $\pm$ 0.003 & -0.003 $\pm$ 0.047 $\pm$ 0.010\\
0.07 $<x_{\mathrm{B}}\le$ 0.10 & 0.001 $\pm$ 0.036 $\pm$ 0.010 & 0.015 $\pm$ 0.041 $\pm$ 0.012 & 0.013 $\pm$ 0.050 $\pm$ 0.012\\
0.10 $<x_{\mathrm{B}}\le$ 0.15 & 0.052 $\pm$ 0.039 $\pm$ 0.007 & 0.042 $\pm$ 0.043 $\pm$ 0.007 & 0.104 $\pm$ 0.054 $\pm$ 0.005\\
0.15 $<x_{\mathrm{B}}\le$ 0.35 & -0.060 $\pm$ 0.052 $\pm$ 0.004 & -0.088 $\pm$ 0.061 $\pm$ 0.004 & -0.081 $\pm$ 0.075 $\pm$ 0.000\\
\hline \hline
1.00 $<Q^2\le$ 1.50 & 0.002 $\pm$ 0.036 $\pm$ 0.010 & 0.015 $\pm$ 0.044 $\pm$ 0.022 & 0.003 $\pm$ 0.053 $\pm$ 0.024\\
1.50 $<Q^2\le$ 2.30 & -0.005 $\pm$ 0.034 $\pm$ 0.002 & 0.009 $\pm$ 0.040 $\pm$ 0.002 & 0.018 $\pm$ 0.049 $\pm$ 0.004\\
2.30 $<Q^2\le$ 3.50 & 0.012 $\pm$ 0.038 $\pm$ 0.005 & 0.014 $\pm$ 0.042 $\pm$ 0.005 & 0.009 $\pm$ 0.053 $\pm$ 0.006\\
3.50 $<Q^2\le$ 10.00 & 0.014 $\pm$ 0.042 $\pm$ 0.002 & 0.000 $\pm$ 0.047 $\pm$ 0.002 & 0.032 $\pm$ 0.059 $\pm$ 0.000\\
\hline \hline
\end{tabular}
\caption{\label{tab:results}
Table of results for the beam-helicity Fourier amplitudes $A\pm\delta A_{\mathrm{stat.}}\pm\delta A_{\mathrm{syst.}}$ extracted from the three exclusive event samples discussed in 
the text: unresolved sample, unresolved-reference sample, and pure sample. 
The correlations between the $\sin\phi$ and $\sin(2\phi)$ amplitudes are, for each of the samples, below 10\%. 
The average kinematic values for each bin are compiled in table \ref{tab:averagekine}. 
The quantities $-t$ and $Q^2$ are given in units of GeV$^2$. A separate scale uncertainty arising from the measurement of the beam polarization amounts to 1.96\%.
}
\end{center}
\end{table}

\section{Summary}
\label{sec:summary}

Azimuthal amplitudes of the beam-helicity asymmetry measured at {\sc Hermes} in exclusive production of real photons of longitudinally polarized positrons incident on an unpolarized hydrogen target are presented. The asymmetry arises from the deeply virtual Compton scattering process and its interference with the Bethe--Heitler process and is sensitive primarily to GPD $H$. 
Azimuthal amplitudes of this asymmetry are extracted from three exclusive data samples: 
from a pure event sample selected by using information from lepton, photon, and recoil-proton detection 
in a kinematic event fit, 
and two unresolved event samples analyzed by means of a missing-mass technique. 
The pure sample consists almost entirely of $ep\rightarrow ep\gamma$ events selected in a kinematically complete measurement, while the unresolved samples are estimated to contain an average 12-14\% contribution from associated production and also an approximate 3\%
contribution from semi-inclusive production, the latter of which is corrected for.
One of the two unresolved samples serves as reference sample for the pure sample, disentangling the effects of the removal of associated background and of the reduced detector acceptance caused by the employment of the recoil detector.
The recoil detector allows the elimination of this background to a very high extent.
After correcting for the acceptance reduction, the removal of associated background from the data sample alone is shown to increase the magnitude of the leading asymmetry amplitude 
by $0.054\pm0.016$ to $-0.328\pm0.027\;\mathrm{(stat.)}\pm0.045\;\mathrm{(syst.)}$.
The leading asymmetry amplitude obtained from the pure sample is well described by recent fits to previously published data. Consistency is found with a theoretical model when a variant is considered that assumes skewness independence for sea quarks (the model shows no sensitivity to the skewness of valence quarks).

\acknowledgments

We gratefully acknowledge the {\sc Desy} management for its support and the staff
at {\sc Desy} and the collaborating institutions for their significant effort.
This work was supported by 
the Ministry of Economy and the Ministry of Education and Science of Armenia;
the FWO-Flanders and IWT, Belgium;
the Natural Sciences and Engineering Research Council of Canada;
the National Natural Science Foundation of China;
the Alexander von Humboldt Stiftung,
the German Bundesministerium f\"ur Bildung und Forschung (BMBF), and
the Deutsche Forschungsgemeinschaft (DFG);
the Italian Istituto Nazionale di Fisica Nucleare (INFN);
the MEXT, JSPS, and G-COE of Japan;
the Dutch Foundation for Fundamenteel Onderzoek der Materie (FOM);
the Russian Academy of Science and the Russian Federal Agency for 
Science and Innovations;
the Basque Foundation for Science (IKERBASQUE) and the UPV/EHU under program UFI 11/55;
the U.K.~Engineering and Physical Sciences Research Council, 
the Science and Technology Facilities Council,
and the Scottish Universities Physics Alliance;
the U.S.~Department of Energy (DOE) and the National Science Foundation (NSF);
as well as the European Community Research Infrastructure Integrating Activity
under the FP7 "Study of strongly interacting matter (HadronPhysics2, Grant
Agreement number 227431)".



\bibliographystyle{natbib}

\begin{thebibliography}{10}
\bibitem{Mueller}
D. M\"uller et al., \emph{Wave Functions, Evolution Equations and Evolution Kernels from Light-Ray Operators of QCD}, Fortschr. Phys. {\bf 42} (1994) 101 [hep-ph/9812448]. 

\bibitem{Radyushkin}
A.~V. Radyushkin, \emph{Scaling Limit of Deeply Virtual Compton Scattering}, Phys. Lett. B \textbf{380} (1996) 417 [hep-ph/9604317].

\bibitem{Ji}
X.~Ji, \emph{Deeply virtual Compton scattering}, Phys. Rev. D \textbf{55} (1997) 7114 [hep-ph/9609381].

\bibitem{Ji_2}
X.~Ji, \emph{Gauge-Invariant Decomposition of Nucleon Spin}, Phys. Rev. Lett. {\bf 78} (1997) 610 [hep-ph/9603249].

\bibitem{Burkardt}
M.~Burkardt, \emph{Impact Parameter Dependent Parton Distributions and Off-Forward Parton Distributions for $\zeta\to 0$}, Phys. Rev. D \textbf{62} (2000) 071503 [hep-ph/0005108].
\\Erratum-ibid. Phys. Rev. D \textbf{66} (2002) 119903. 

\bibitem{Air01}
H{\sc ermes} Collaboration, A. Airapetian et al., \emph{Measurement of the Beam-Spin Azimuthal Asymmetry Associated with Deeply-Virtual Compton Scattering}, Phys. Rev. Lett. \textbf{87} (2001) 182001 [hep-ex/0106068].

\bibitem{Air06}
H{\sc ermes} Collaboration, A. Airapetian et al., \emph{The beam-charge azimuthal asymmetry and deeply virtual Compton scattering}, Phys. Rev. D \textbf{75} (2007) 011103R [hep-ex/0605108].

\bibitem{PublicationDraft69}
H{\sc ermes} Collaboration, A. Airapetian et al., \emph{Separation of contributions from deeply virtual Compton scattering and its interference with the Bethe--Heitler process in measurements on a hydrogen target}, JHEP {\bf 11} (2009) 083 [arXiv:0909.3587].

\bibitem{PublicationDraft90}
H{\sc ermes} Collaboration, A. Airapetian et al., \emph{Beam-helicity and beam-charge asymmetries associated with deeply virtual Compton scattering on the unpolarised proton}, accepted by JHEP [arXiv:1203.6287].

\bibitem{h101}
H1 Collaboration, C. Adloff et al., \emph{Measurement of deeply virtual Compton scattering at HERA}, Phys. Lett. B {\bf 517} (2001) 47 [hep-ex/0107005].
 
\bibitem{h105}
H1 Collaboration, A. Aktas et al., \emph{Measurement of deeply virtual Compton scattering at HERA}, Eur. Phys. J. C  {\bf 44} (2005) 1 [hep-ex/0505061].

\bibitem{h107}
H1 Collaboration, F. D. Aaron et al., \emph{Measurement of deeply virtual Compton scattering and its $t$-dependence at HERA}, Phys. Lett. B {\bf 659} (2007) 796 [arXiv:0709.4114].

\bibitem{h109}
H1 Collaboration, F. D. Aaron et al., \emph{Deeply virtual Compton scattering and its Beam Charge Asymmetry in $e^{\pm}$ Collisions at HERA}, Phys. Lett. B {\bf 681} (2009) 391 [arXiv:0907.5289].

\bibitem{zeu03}
Z{\sc eus} Collaboration, S. Chekanov et al., \emph{Measurement of deeply virtual Compton scattering at HERA}, Phys. Lett. B {\bf 573} (2003) 46 [hep-ex/0305028].

\bibitem{zeu08}
Z{\sc eus} Collaboration, S. Chekanov et al., \emph{A measurement of the $Q^2$, $W$ and $t$ dependences of deeply virtual Compton scattering at HERA}, JHEP {\bf 05} (2009) 108 [arXiv:0812.2517].

\bibitem{Cam06}
Jefferson Lab Hall A Collaboration,  C. Camacho et al., \emph{Scaling Tests of the Cross Section for Deeply Virtual Compton Scattering}, Phys. Rev. Lett. \textbf{97} (2006) 262002 [arXiv:nucl-ex/0607029].

\bibitem{Ste01}
C{\sc las} Collaboration, S. Stepanyan et al., \emph{Observation of exclusive DVCS in polarized electron beam asymmetry measurements}, Phys. Rev. Lett. {\bf 87} (2001) 182002 [hep-ex/0107043].

\bibitem{Gir08}
C{\sc las} Collaboration, F. X. Girod et al., \emph{Measurement of Deeply Virtual Compton Scattering Beam-Spin Asymmetries}, Phys. Rev. Lett. \textbf{100} (2008) 162002 [arXiv:0711.4805].

\bibitem{Gav09}
C{\sc las} Collaboration, G. Gavalian et al., \emph{Beam Spin Asymmetries in DVCS with CLAS at 4.8 GeV}, Phys. Rev. C {\bf 80} (2009) 035206 [arXiv:0812.2950].

\bibitem{DVCS2}
A.V. Belitsky, D. M\"uller and A. Kirchner, \emph{Theory of Deeply Virtual Compton Scattering on the nucleon}, Nucl. Phys. B \textbf{629} (2002) 323 [hep-ph/0112108].

\bibitem{Tre04}
A. Bacchetta et al., \emph{Single-spin asymmetries: The Trento conventions}, Phys. Rev. D \textbf{70} (2004) 117504 [hep-ph/0410050].

\bibitem{RecoilTechnicalPaper}
 A.~Airapetian \etal, \emph{The HERMES Recoil Detector}, to be submitted to JINST.

\bibitem{hermes:spectrometer}
HERMES Collaboration, K. Ackerstaff et al., \emph{The HERMES Spectrometer}, Nucl. Instrum. Meth. A {\bf 417} (1998)~230 [hep-ex/9806008].

\bibitem{Sokolov+:1964}
A.~Sokolov and I.~Ternov, \emph{On polarization and spin effects in the theory of synchrotron radiation}, Sov.\ Phys.\ Doklady\ {\bf 8} (1964) 1203.

\bibitem{Buon:1986}
J.~Buon and K.~Steffen, \emph{HERA Variable Energy 'Mini' Spin Rotator and head-on ep Collision scheme with Choice of Electron Helicity}, Nucl.\ Instrum.\ and Meth. A {\bf 245} (1986) 248.

\bibitem{TPOL:1994}
D.P.~Barber \etal, \emph{High Spin Polarisation at the HERA Electron Storage Ring}, Nucl.\ Instrum.\ and Meth. A {\bf 338} (1994) 166.

\bibitem{LPOL:2002}
M.~Beckmann \etal, \emph{The Longitudinal Polarimeter at HERA}, Nucl.\ Instrum.\ and Meth. A {\bf 479} (2002) 334 [arXiv:physics/0009047].

\bibitem{HERApol2012}
B.~Sobloher, R.~Fabbri, T.~Behnke, J.~Olsson, D.~Pitzl, S.~Schmitt, J.~Tomaszewska, \emph{Polarisation at HERA - Reanalysis of the HERA II Polarimeter Data}, [arXiv:1201.2894].

\bibitem{Xianguo}
X.~Lu, \emph{The HERMES Recoil Detector: Particle Identification and Determination of Detector Efficiency of the Scintillating Fiber Tracker}, Master Thesis, Universit\"at Hamburg, Germany, September 2009, DESY-THESIS-2009-043.

\bibitem{FranksPhD}
F.~Ellinghaus, \emph{Beam-Charge and Beam-Spin Azimuthal Asymmetries in Deeply-Virtual Compton Scattering}, Ph.D. thesis, Humboldt Universit\"at zu Berlin, Germany, November 2003, DESY-THESIS-2004-005. 

\bibitem{lepto}
G.~Ingelman, A.~Edin, J.~Rathsman, \emph{LEPTO 6.5 - A Monte Carlo Generator for Deep Inelastic Lepton-Nucleon Scattering}, Comput. Phys. Commun. {\bf 101} (1997) 108 [hep-ph/9605286].


\bibitem{gmcDVCS}
V.~A.~Korotkov and W.~D.~Nowak, \emph{Future Measurements of Deeply Virtual Compton Scattering at HERMES}, Eur. Phys. J. C {\bf 23} (2002) 455 [hep-ph/018077].

\bibitem{motsai}
L.~W.~Mo and Y.~S.~Tsai, \emph{Radiative Corrections to Elastic and Inelastic ep and $\mu$p Scattering}, Rev.~Mod.~Phys. {\bf 41} (1969) 205.

\bibitem{BernhardsPhD}
B.~Krau\ss, \emph{Deeply Virtual Compton Scattering and the HERMES-Recoil-Detector}, Ph.D. thesis, Friedrich-Alexander Universit\"at Erlangen-N\"urnberg, Germany, February 2005, DESY-THESIS-2005-007.

\bibitem{Bra76}
F.~W.~Brasse et al., \emph{Parametrization of the $Q^2$ dependence of $\gamma_V p$ total cross sections in the resonance region}, Nucl. Phys. B {\bf 110} (1976) 413.

\bibitem{radgen}
I.~Akushevich, H.~B\"ottcher, and D.~Ryckbosch, \emph{RADGEN 1.0. Monte Carlo Generator for Radiative Events in DIS on Polarized and Unpolarized Targets}, [hep-ph/9906408].

\bibitem{PublicationDraft68}
HERMES Collaboration, A. Airapetian et al., \emph{Measurement of azimuthal asymmetries with respect to both beam charge and transverse target polarization in exclusive electroproduction of real photons}, JHEP \textbf{06} (2008) 066 [arXiv:0802.2499].

\bibitem{MML}
R. Barlow, \emph{Extended maximum likelihood}, Nucl. Instrum. Meth. A {\bf 297} (1990) 496.


\bibitem{PhDZhenyu}
Z.~Ye, \emph{Transverse target-spin asymmetry associated with deeply virtual Compton scattering on
the proton and a resulting model-dependent constraint on the total angular momentum of
quarks in the nucleon}, Ph.D. thesis, Universit\"at Hamburg, Germany, December 2006, DESY-THESIS-2007-005.

\bibitem{PhDArne}
A.~Vandenbroucke, \emph{Exclusive $\pi^0$ Production at HERMES: Detection - Simulation - Analysis}, Ph.D. thesis, Universiteit Gent, Belgium, November 2006, DESY-THESIS-2007-003.

\bibitem{exclpion}
H{\sc ermes} Collaboration, A. Airapetian et al., \emph{Cross sections for hard exclusive electroproduction of $\pi^+$ mesons on a hydrogen target}, Phys. Lett. B \textbf{659} (2008) 486 [arXiv:0707.0222].

\bibitem{Vdhcode}
M.~Vanderhaeghen, P.A.M.~Guichon, and M.~Guidal, \emph{Computer code for the calculation of DVCS and BH processes in the reaction $e p \rightarrow e p \gamma$}, private communication, 2007.

\bibitem{Vanderhaeghen:1999xj} 
M.~Vanderhaeghen, P.A.M.~Guichon, and M.~Guidal, \emph{Deeply virtual electroproduction of photons and mesons on the nucleon: Leading order amplitudes and power corrections}, Phys.\ Rev.\ D {\bf 60} (1999) 094017 [hep-ph/9905372].

\bibitem{Kum09}
K. Kumeri\v{c}ki and D. M\"uller, \emph{Deeply virtual Compton scattering at small $x_B$ and the access to the GPD H}, Nucl. Phys. B {\bf 841} (2010) 1 [arXiv:0904.0458].

\bibitem{AssoAsy}
P.~A.M.~Guichon, L.~Moss\'e, and M.~Vanderhaeghen, \emph{Pion production in deeply virtual Compton scattering}, Phys.\ Rev.\ D {\bf 68} (2003) 034018 [hep-ph/0305231].

\end{thebibliography}

\end{document}